\documentclass[preprint,8pt]{elsarticle}

\usepackage{amssymb}
\usepackage{graphicx}
\usepackage{amssymb}
\usepackage{graphicx}
\usepackage{amsfonts}
\usepackage{geometry}
\usepackage{epsfig}
\usepackage{CJK}
\usepackage{amsmath}
\usepackage{epstopdf}
\usepackage{mathrsfs}
\usepackage{bbding}
\usepackage{cases}
\usepackage{subfigure}
\usepackage{float}
\usepackage{lineno,hyperref}
\usepackage{lineno,hyperref}
\usepackage{lipsum}
\usepackage{amssymb}
\usepackage{graphicx}
\usepackage{amsfonts}
\usepackage{geometry}
\usepackage{epsfig}
\usepackage{CJK}
\usepackage{amsmath}
\usepackage{graphicx}
\usepackage{epstopdf}
\usepackage{mathrsfs}
\usepackage{bbding}
\usepackage{cases}
\usepackage{subfigure}
\usepackage{float}
\usepackage{lineno,hyperref}
\modulolinenumbers[5]

\journal {}

\begin{document}

\begin{frontmatter}



\title{Multi-dark soliton solutions for the multi-component coupled Maccari system }

\author[]{Tao Xu$^{1}$}
\author[]{Yong Chen$^{1,2}$\corref{cor1}}
\cortext[cor1]{Corresponding author.}\ead{ychen@sei.ecnu.edu.cn}
\address{1 Institute of Computer Theory, East China Normal University, Shanghai, 200062, China\\
2 Department of Physics, Zhejiang Normal University, Jinhua 321004, China}

\begin{abstract}
Based on the KP hierarchy reduction method, the general multi-dark soliton solutions in Gram type determinant forms for the multi-component coupled Maccari system are constructed. Especially, the three-component coupled Maccari system comprised of two-component short waves and single-component long waves are discussed in detail. Besides, the dynamics of one and two dark-dark solitons are analyzed. It is shown that the collisions of two dark-dark solitons are elastic by asymptotic analysis. Additionally, the two dark-dark solitons bound states are studied through two different cases (stationary and moving cases). The bound states can exist up to arbitrary order in the stationary case, however, only two-soliton bound state exists in the moving case. Besides, the oblique stationary bound state can be generated for all possible combinations of nonlinearity coefficients consisted of positive, negative and mixed cases. Nevertheless, the parallel stationary and the moving bound states are only possible when nonlinearity coefficients take opposite signs.
\end{abstract}


\end{frontmatter}



\section{Introduction}
The study for multi-component coupled nonlinear systems has grown popular rapidly in recent years, and a lot of novel and interesting physical phenomenons can be generated by the nonlinear interaction of multiple waves \cite{8-1,8-2,8-3,8-4,8-5}.
 Additionally, a variety of complex systems usually involve more than one component, such as nonlinear optical fibers and Bose-Einstein condensates, etc.. It is greatly necessary to extend the corresponding researches to multi-component systems \cite{8-6,8-7,8-8}. There always exist coupled effects that cross-phase modulation in the multi-component ones, and the corresponding various solutions can not be correlated by Galileo transformation \cite{8-9}. Compared to the single-component systems, affluent exact solutions may be appeared in multi-component systems. In \cite{8-10}, some anomalous Peregrine solitons were constructed for the coupled Fokas-Lenells equations. $N$th-order vector rational and semi-rational rogue waves for the three-component nonlinear Schr\"{o}dinger (NLS) equations were given by Darboux transformation \cite{8-11}. Utilizing the KP hierarchy reduction technique, hybrid solutions were exhibated in the Mel'nikov system \cite{8-12}. Furthermore, the interactional solutions including higher-order RWs interacting with multi-soliton (or multi-breather) were constructed in different multi-component systems \cite{8-13,8-14}.

As one of the hot topics in the nonlinear science, dark solitons  have been reported in many documents \cite{8-15,8-16,8-17,8-18}. There have been many different methods for constructing the dark soliton, such as the binary Darboux transformation \cite{8-15,8-19,8-20}, the algebraic-geometry reduction method \cite{8-17}, the dressing-Hirota method \cite{8-18} and the KP hierarchy reduction method \cite{8-16,8-21,8-22}. The main goal of this work is to construct multi-dark soliton solutions for the multi-component coupled Maccari system through the KP hierarchy reduction method. In 1983, the Kyoto school first developed the KP hierarchy reduction
method \cite{8-23}, which have been widely applied to generate dark soliton \cite{8-16,8-21,8-22}, bright-dark mixed soliton \cite{8-24,8-25,8-26}, general rogue wave solutions \cite{8-27,8-28,8-29} and interactional solutions \cite{8-12,8-30} for a lot of integrable nonlinear systems.

In this paper, we focus on the multi-dark soliton solutions for the following three-component coupled Maccari system and its $(N+1)$-component generalization

\begin{eqnarray}
&&i\phi^{(1)}_t+\phi^{(1)}_{xx}+v\phi^{(1)}=0,\label{xt-8-1}\\
&&i\phi^{(2)}_t+\phi^{(2)}_{xx}+v\phi^{(2)}=0,\label{xt-8-3}\\
&&v_y=(\sigma_1\phi^{(1)}\phi^{(1)*}+\sigma_2\phi^{(2)}\phi^{(2)*})_x,\label{xt-8-2}
\end{eqnarray}
where $\sigma_k=\pm1 (k=1,2)$, $\phi^{(1)}\equiv\phi^{(1)}(x,y,t)$ and $\phi^{(2)}\equiv\phi^{(2)}(x,y,t)$ are the complex short waves amplitudes and $v\equiv v(x,y,t)$  is the real long wave amplitude. Besides, the symbol $^{*}$ denotes complex
conjugation and the subscripted variables $x$, $y$ and $t$ in Eqs. (\ref{xt-8-1})-(\ref{xt-8-2}) are the corresponding partial differentiations. The above coupled Maccari system (\ref{xt-8-1})-(\ref{xt-8-2}) can be generalized to a $(N+1)$-component coupled system including  $N$
short wave components and one long wave component, which can be  written as follows

\begin{eqnarray}
&&i\phi^{(l)}_{t}+\phi_{xx}+v\phi^{(l)}=0,~l=1,2,\dots,N,\label{xt-8-4}\\
&&v_y=\sum_{l=1}^{N}(\sigma_{l}\phi^{(l)}\phi^{(l)*})_x,\label{xt-8-5}
\end{eqnarray}
 where $\sigma_l=\pm1$, $\phi^{(l)}\equiv\phi^{(l)}(x,y,t)~(l=1,2,\dots,N)$ denote $N$ different complex short waves amplitudes and $v\equiv v(x,y,t)$ is the real long wave amplitude.

 Through  a special reduction method, the coupled system (\ref{xt-8-1})-(\ref{xt-8-2}) were first derived by Maccari \cite{8-31}. Under some special transformations, the above mentioned three-component coupled Maccari system can be reduced to our known nonlinear models. When $y=x$, the coupled system (\ref{xt-8-1})-(\ref{xt-8-2}) becomes two-component coupled NLS equations \cite{8-31-1}. Choosing $\phi^{(1)}=\phi^{(2)*}$, the Maccari system is reduced to a $(2+1)$-dimensional extension of the NLS equation \cite{8-31-2}. If $y=t$, we can get the coupled long-wave resonance equations \cite{8-31-3}. The Maccari system can describe some motions of isolated waves which are localized in small parts of space in many field, such as nonlinear optics, hydrodynamics, plasma physics \cite{8-32}.

 There have been many articles reported on the coupled Maccari system. Utilizing bilinear method, the solitoff solutions of the Maccari system (\ref{xt-8-1})-(\ref{xt-8-2}) were given in \cite{8-33}. The Painlev\'{e} integrability and some special types of the localized excitations  were constructed  in \cite{8-34}. Some types of doubly periodic propagating wave patterns were generated by the variable separation approach \cite{8-35}. In \cite{8-36}, dromion solutions for the Maccari system were generated. Different types of rational solutions including multi-rogue waves, lump solitons and interactions between the two were all given in the Maccari system by bilinear method \cite{8-32}. Additionally, the bright solitons \cite{8-37} and bright-dark mixed solitons \cite {8-26,8-38} for the Maccari system were all studied by different methods. To the best of our knowledge, multi-dark soliton solutions for the coupled Maccari system (\ref{xt-8-1})-(\ref{xt-8-2}) and the corresponding multi-component generalization (\ref{xt-8-4})-(\ref{xt-8-5}) have never been reported up to now. Here, we will construct the multi-dark soliton solutions for the Maccari system by the KP hierarchy reduction method.

 In the present paper, the $N$-dark soliton solutions in Gram type determinant are generated through the KP hierarchy reduction method. Furthermore, the three-component coupled case (\ref{xt-8-1})-(\ref{xt-8-2}) are discussed in detail for an example. Some dynamics and figures of one and two dark-dark solitons are given. Utilizing the asymptotic analysis technique, the collisions between the two solitons are proved to be elastic. It can be shown that the two dark-dark solitons bound states possess two cases consisted of  stationary and moving bound states. The bound states can exist up to arbitrary order in the stationary case, however, only two-soliton bound state exists in the moving case. Moreover, the oblique stationary bound state can be generated for all possible combinations of nonlinearity coefficients consisted of positive, negative and mixed cases. Nevertheless, the parallel stationary and the moving bound states are only possible when nonlinearity coefficients take opposite signs.

The paper is organized as follows. In section 2, the formulae of $N$-dark-dark soliton solutions for the three-component coupled Maccari system (\ref{xt-8-1})-(\ref{xt-8-2}) are given and proved. In section 3, some dynamics of one- and two-dark-dark solitons are discussed in detail. In section 4, the dark-dark soliton bound states including stationary and moving cases are exhibited. In section 5, the uniform formulae of $N$-dark solitons for the multi-component generalization (\ref{xt-8-4})-(\ref{xt-8-5}) are constructed. The last section includes some conclusions and discussions.

\section{N-dark-dark soliton solutions for the three-component coupled Maccari system }
Through the dependent variable transformation
\begin{equation}
\phi^{(1)}=k_1e^{i\theta_1}\frac{h^{(1)}}{f},\quad \phi^{(2)}=k_2 e^{i\theta_2}\frac{h^{(2)}}{f},\quad v=2(lnf)_{xx},
\end{equation}
the three-component coupled Maccari system (\ref{xt-8-1})-(\ref{xt-8-2}) can be transformed to the following bilinear forms
\begin{eqnarray}
&&[D_x^2+i(D_t+2\alpha_1D_x)] h^{(1)}\cdot f=0,\label{xt-8-12}\\
&&[D_x^2+i(D_t+2\alpha_2D_x)] h^{(2)} \cdot f=0,\label{xt-8-13}\\
&&D_xD_y f\cdot f=\sigma_1 k_1^2(h^{(1)}h^{(1)*}{-}f^2)+\sigma_2 k_2^2(h^{(2)}h^{(2)*}{-}f^2),\label{xt-8-14}
\end{eqnarray}
where $k_1$ and $k_2$ are arbitrary real constants, $f$, $h^{(1)}$ and $h^{(2)}$  all are the functions of $x$, $y$ and $t$, besides, $f$ is a real function, $h^{(1)}$ and $h^{(2)}$ are two complex functions. Meanwhile, $\theta_i=\alpha_ix-\alpha_i^2 t+\beta_i(y)~(i=1,2)$, where $\alpha_i$ are arbitrary real constants $(\alpha_1\neq\alpha_2)$ and $\beta_i(y)$ are  arbitrary real functions of $y$.  The operator $D$ denotes the  Hirota¡¯s bilinear differential operator, which can be defined as follows
\begin{equation}
D_x^{l}D_y^{m}D_t^{n} f(x,y,t)\cdot g(x,y,t)=(\frac{\partial}{\partial_x}-\frac{\partial}{\partial_{x'}})^{l}(\frac{\partial}{\partial_y}-\frac{\partial}{\partial_{ y'}})^{m}(\frac{\partial}{\partial_{t}}-\frac{\partial}{\partial_{t'}})^{n}f(x,y,t)g(x',y',t')|_{x=x',y=y',t=t'}.
\end{equation}

From the KP theory, the nonlinear models under investigation can be seen as some special reductions of the integrable  equations in the KP hierarchy \cite{8-7,8-16}. Then the solutions that bright, dark and rational solutions for some relative nonlinear systems can be derived from the $\tau$ function in the KP hierarchy under  the appropriate reductions. Here, the multi-dark soliton solutions for the Maccari system (\ref{xt-8-1})-(\ref{xt-8-2}) can be generated from the reduction of single KP hierarchy consisted of two copies of shifted singular points.

\textbf{Lemma 1.}
Considering the following bilinear equations in the KP hierarchy \cite{8-16,8-21,8-22}
\begin{eqnarray}
&&(D_{x_1}^2+2aD_{x_1}-D_{x_2})\tau(k{+}1,l)\cdot\tau(k,l)=0,\label{xt-8-6}\\
&&(\frac{1}{2}D_{x_1}D_{x_{-1}}-1)\tau(k,l)\cdot\tau(k,l)=-\tau(k{+}1,l)\tau(k{-}1,l),\label{xt-8-6-1}\\
&&(D_{x_1}^2+2bD_{x_1}-D_{x_2})\tau(k,l{+}1)\cdot\tau(k,l)=0,\label{xt-8-6-2}\\
&&(\frac{1}{2}D_{x_1}D_{y_{-1}}-1)\tau(k,l)\cdot \tau(k,l)=-\tau(k,l{+}1)\tau(k,l{-}1),\label{xt-8-7}
\end{eqnarray}
where $k$ and $l$ are arbitrary integers, $a$ and $b$ are two complex constants. The above bilinear equations (\ref{xt-8-6})-(\ref{xt-8-7}) have the Gram determinant solutions
\begin{equation}
\tau(k,l)=|m_{(i,j)}(k,l)|_{1\leq i,j\leq N},
\end{equation}
here, the entries of the determinant $\tau(k,l)$ are given as
\begin{eqnarray}
&&m_{i,j}(k,l)=c_{ij}+\displaystyle{\int \varphi_i(k,l)\psi_j(k,l)d{x_1}},\\
&&\varphi_i(k,l)=(p_i-a)^k(p_i-b)^le^{\eta_i},\\
&&\psi_j(k,l)=(-\frac{1}{q_j+a})^k(-\frac{1}{q_j+b})^{l}e^{\zeta_j},
\end{eqnarray}
with
\begin{eqnarray*}
&&\eta_i=\frac{1}{p_i-a}x_{-1}+\frac{1}{p_i-b}y_{-1}+p_ix_1+p_i^2x_2+\eta_{i0},\\
&&\zeta_j=\frac{1}{q_j+a}x_{-1}+\frac{1}{q_j+b}y_{-1}+q_jx_1-q_j^2x_2+\zeta_{j0},
\end{eqnarray*}
where $c_{ij}$,$p_i$,$q_j$, $\eta_{i0}$ and $\zeta_{j0}$ are complex constants.

We should point out that, Eqs. (\ref{xt-8-6}) and (\ref{xt-8-6-2}) are the lowest-degree bilinear equation in the 1st modified KP hierarchy \cite{8-23}, Eqs. (\ref{xt-8-6-1}) and (\ref{xt-8-7}) are the bilinear equation for the two-dimensional Toda lattice \cite{8-23,8-40-1}. The validity of Lemma 1 have been proved in Eq.(6) in the reference \cite{8-16} and we omit the processes of proof here.

In order to get the $N$-dark-dark soliton solutions of the Maccari system (\ref{xt-8-1})-(\ref{xt-8-2}), we should consider the reduction of the bilinear equations in Lemma 1. Choosing that $a$,~$b$ and $x_2$ are pure  imaginary, $x_1$,~$x_{-1}$,~$y_{-1}$ are real, and fixing the following relations $a=i\alpha_1$,~$b=i\alpha_2$,~$q_j=p_j^{*}$,~$\zeta_{j0}=\eta_{j0}^{*}$,~$c_{ji}=c_{ij}=\delta_{ij}$~(~$\delta_{ij}$~is the Kronecker symbol),the following equalities can be given
\begin{equation}
\zeta_j=\eta_j^{*},\quad \tau(k,l)=\tau^{*}(-k,-l).
\end{equation}
Through defining $f$,~$h^{(1)}$,~$h^{(2)}$ as
\begin{equation}
f=\tau(0,0),\quad h^{(1)}=\tau(1,0),\quad h^{(2)}=\tau(0,1),\quad h^{(1)*}=\tau(-1,0),\quad h^{(2)*}=\tau(0,-1), \label{xt-8-16}
\end{equation}
the bilinear equations (\ref{xt-8-6})-(\ref{xt-8-7}) can be rewritten as
\begin{eqnarray}
&&(D_{x_1}^2+2i\alpha_1D_{x_1}-D_{x_2})h^{(1)}\cdot f=0,\label{xt-8-8}\\
&&(\frac{1}{2}D_{x_1}D_{x_{-1}}-1)f\cdot f=-h^{(1)}h^{(1)*},\label{xt-8-9}\\
&&(D_{x_1}^2+2i\alpha_2D_{x_1}-D_{x_2})h^{(2)}\cdot f=0,\label{xt-8-10}\\
&&(\frac{1}{2}D_{x_1}D_{y_{-1}}-1)f\cdot f=-h^{(2)}h^{(2)*}.\label{xt-8-11}
\end{eqnarray}
 If we  introduce these independent variable transformations
\begin{equation}
x_1=x,\quad x_2=it,\quad x_{-1}=-\frac{1}{2}\sigma_1k_1^2y,\quad y_{-1}=-\frac{1}{2}\sigma_2k_2^2y,\label{xt-8-15}
\end{equation}
namely
\begin{equation}
\partial_x=\partial_{x_1},\quad \partial_t=i\partial_{x_2},\quad \partial_y=-\frac{1}{2}\sigma_1k_1^2\partial_{x_{-1}}-\frac{1}{2}\sigma_2k_2^2\partial_{y_{-1}},
\end{equation}
the bilinear  Eqs.(\ref{xt-8-8})-(\ref{xt-8-11}) can be directly calculate to recast to  Eqs. (\ref{xt-8-12})-(\ref{xt-8-14}). Considering both the $\tau$ functions in Eq. (\ref{xt-8-16}) and the variable transformations (\ref{xt-8-15}), one can directly get the $N$-dark-dark soliton solution in  Gram type determinant
forms.

\textbf{Theorem 1.} \label{theorem1}
The $N$-dark-dark soliton solutions of the three-component coupled Maccari system (\ref{xt-8-1})-(\ref{xt-8-2}) can be written as
\begin{equation}
\phi^{(1)}=k_1e^{i\theta_1}\frac{h^{(1)}}{f},\quad \phi^{(2)}=k_2 e^{i\theta_2}\frac{h^{(2)}}{f},\quad v=2(lnf)_{xx}, \label{xt-8-17}
\end{equation}
where $f$,~$h^{(1)}$ and $h^{(2)}$ are given in the following Gram determinants
\begin{eqnarray}
&&f=\left|\delta_{ij}+\frac{1}{p_i+p_j^{*}}e^{\xi_i+\xi_j^{*}}\right|_{N\times N},\\
&&h^{(1)}=\left|\delta_{ij}+(-\frac{p_i-i\alpha_1}{p_j^{*}+i\alpha_1})\frac{1}
{p_i+p_j^{*}}e^{\xi_i+\xi_j^{*}}\right|_{N\times N},\\
&&h^{(2)}=\left|\delta_{ij}+(-\frac{p_i-i\alpha_2}{p_j^{*}+i\alpha_2})\frac{1}
{p_i+p_j^{*}}e^{\xi_i+\xi_j^{*}}\right|_{N\times N},\label{xt-8-18}
\end{eqnarray}
with
\begin{eqnarray*}
&&\theta_1=\alpha_1x-\alpha_1^2 t+\beta_1(y),\quad \theta_2=\alpha_2x-\alpha_2^2 t+\beta_2(y),\\
&&\xi_i=p_ix+ip_i^2t-\frac{1}{2}(\frac{\sigma_1k_1^2}{p_i-i\alpha_1}+\frac{\sigma_2k_2^2}{p_i-i\alpha_2})y
+\xi_{i0},~(i=1,2,\cdots,N),
\end{eqnarray*}
where $\beta_1(y)$ and $\beta_2(y)$ are all arbitrary real functions of $y$, $p_i$ and $\xi_{i0}$ are complex constants,~$\delta_{ij}$ is the Kronecker symbol

\section{Dynamics of the $N$-dark-dark soliton solutions}  
\subsection{One dark-dark solitons}
In this paper, the long wave component $v$ always possess bright soliton and we consider to construct dark solitons in the short wave components. In order to get one dark-dark soliton for the three-component coupled Maccari system (\ref{xt-8-1})-(\ref{xt-8-2}), we choose $N=1$ in the formulae (\ref{xt-8-17})-(\ref{xt-8-18}). The concrete expressions in Gram type determinant
forms can be given as
\begin{eqnarray}
&& f_1=1+\frac{1}{p_1+p_1^{*}}e^{\xi_1+\xi_1^{*}},\\
&&h_1^{(1)}=1-\frac{p_1-i\alpha_1}{p_1^{*}+i\alpha_1}\frac{1}{p_1+p_1^{*}}e^{\xi_1+\xi_1^{*}},\\
&&h_1^{(2)}=1-\frac{p_1-i\alpha_2}{p_1^{*}+i\alpha_2}\frac{1}{p_1+p_1^{*}}e^{\xi_1+\xi_1^{*}},
\end{eqnarray}
then substituting the above expressions into the dependent variable transformation (\ref{xt-8-17}), the one dark-dark soliton solutions can read as
\begin{eqnarray}
&&\phi^{(1)}=\frac{k_1}{2}e^{i\theta_1}[1+H_1^{(1)}+(H_1^{(1)}-1)tanh(\frac{\xi_1+\xi_1^{*}+\Theta_1}{2})],\label{xt-8-19}\\
&&\phi^{(2)}=\frac{k_2}{2}e^{i\theta_2}[1+H_1^{(2)}+(H_1^{(2)}-1)tanh(\frac{\xi_1+\xi_1^{*}+\Theta_1}{2})],\\
&&v=\frac{(p_1+p_1^{*})^2}{2}sech^{2}(\frac{\xi_1+\xi_1^{*}+\Theta_1}{2}),\label{xt-8-20}
\end{eqnarray}
where
\begin{eqnarray*}
&&e^{\Theta_1}=\frac{1}{p_1+p_1^{*}}=\frac{1}{2m_1},\quad H_1^{(1)}=-\frac{p_1-i\alpha_1}{p_1^{*}+i\alpha_1}=-\frac{m_1+i(n_1-\alpha_1)}{m_1-i(n_1-\alpha_1)},\\ &&H_1^{(2)}=-\frac{p_1-i\alpha_2}{p_1^{*}+i\alpha_2}=-\frac{m_1+i(n_1-\alpha_2)}{m_1-i(n_1-\alpha_2)},\quad \xi_1+\xi_1^{*}=2m_1x-4m_1n_1t\\
&&-(\frac{\sigma_1m_1k_1^2}{m_1^2+(n_1-\alpha_1)^2}+
\frac{\sigma_2m_1k_2^2}{m_1^2+(n_1-\alpha_2)^2})y+2\xi_{10R},
\end{eqnarray*}
here, $p_1=m_1+in_1$, $\xi_{10}=\xi_{10R}+i\xi_{10I}$ and $m_1$,~$n_1$,~$\xi_{10R}$,~$\xi_{10I}$ are all real constants. Choosing $m_1>0$, we can get the nonsingular one dark-dark soliton solutions.

From Eqs. (\ref{xt-8-19})-(\ref{xt-8-20}), we can directly calculate that the two short wave components' intensity functions $|\phi^{(1)}|$,~$|\phi^{(2)}|$ and the long wave component's amplitude $v$ move at velocity $2n_1$ along the $x$-direction, and at $\dfrac{-4n_1}{\frac{\sigma_1k_1^2}{m_1^2+(n_1-\alpha_1)^2}+\frac{\sigma_2k_2^2}{m_1^2+(n_1-\alpha_2)^2}}$ along the $y$-direction. When $x,~y\rightarrow \pm \infty$, we can derived that $|\phi^{(1)}|\rightarrow |k_1|$,~$\phi^{(2)}|\rightarrow |k_2|$ and $v\rightarrow 0$.

The two complex constants $H_1^{(1)}$ and $H_1^{(2)}$ can be written as the following forms $H_1^{(1)}=e^{2i\psi_1^{(1)}}$ and $H_1^{(2)}=e^{2i\psi_1^{(2)}}$~($-\frac{\pi}{2}\leq\psi_1^{(1)},~\psi_1^{(2)}\leq \frac{\pi}{2}$). With $x$ and $y$ varying from $-\infty$ to $+\infty$, the phase shifts of the two short wave components $\phi^{(1)}$ and $\phi^{(2)}$ are $2\psi_1^{(1)}$ and $\psi_1^{(2)}$ respectively, but the phase shift of the long wave
component $v$  equals zero. Choosing $\xi_1+\xi_1^{*}+\Theta_1=0$, we can directly calculate that the intensities of the center for the three solitons as $|\phi^{(1)}|_{center}=|k_1|cos\psi_1^{(1)}$,~$|\phi^{(2)}|_{center}=|k_2|cos\psi_1^{(2)}$ and $v_{center}=2m_1^2$. For the two short wave components, it can be shown that the center intensities $|\phi^{(i)}|_{center}=|k_i|cos\psi_1^{(i)}$ are smaller than the plane background $|k_i|$ and these two solitons are dark solitons. However, the soliton in the long wave component is bright. Additionally, the degrees of the darkness in the dark solitons in the two short wave components are controlled by the corresponding phase shifts $2\psi_1^{(1)}$ and $2\psi_1^{(2)}$.

According to the expressions of $H_1^{(1)}$ and $H_1^{(2)}$, we can discuss the degrees of the darkness for the two dark solitons in two different cases.

\begin{figure}[!ht]
\renewcommand{\figurename}{{Fig.}}
\subfigure[]{\includegraphics[height=0.3\textwidth]{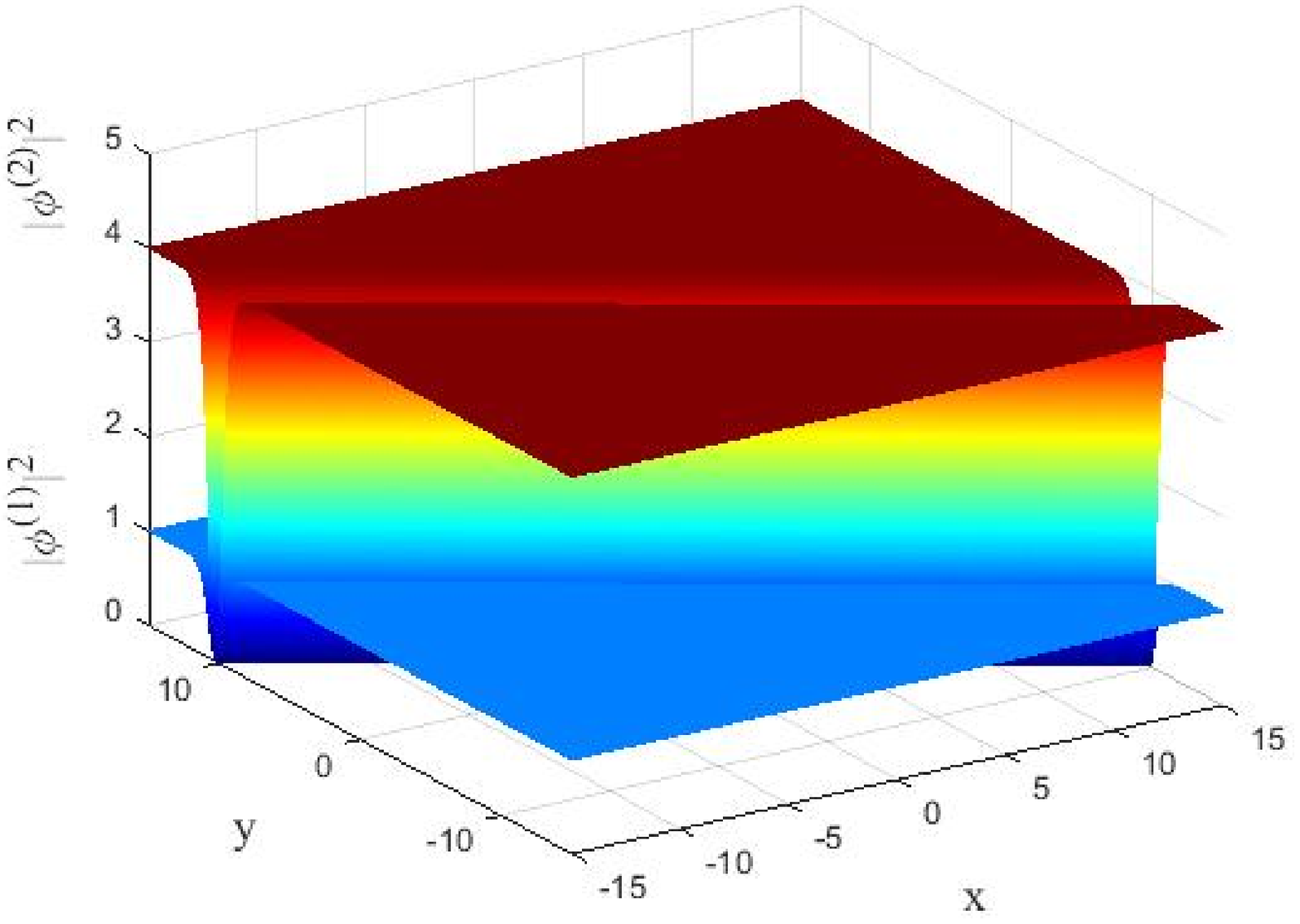}}
\centering
\subfigure[]{\includegraphics[height=0.3\textwidth]{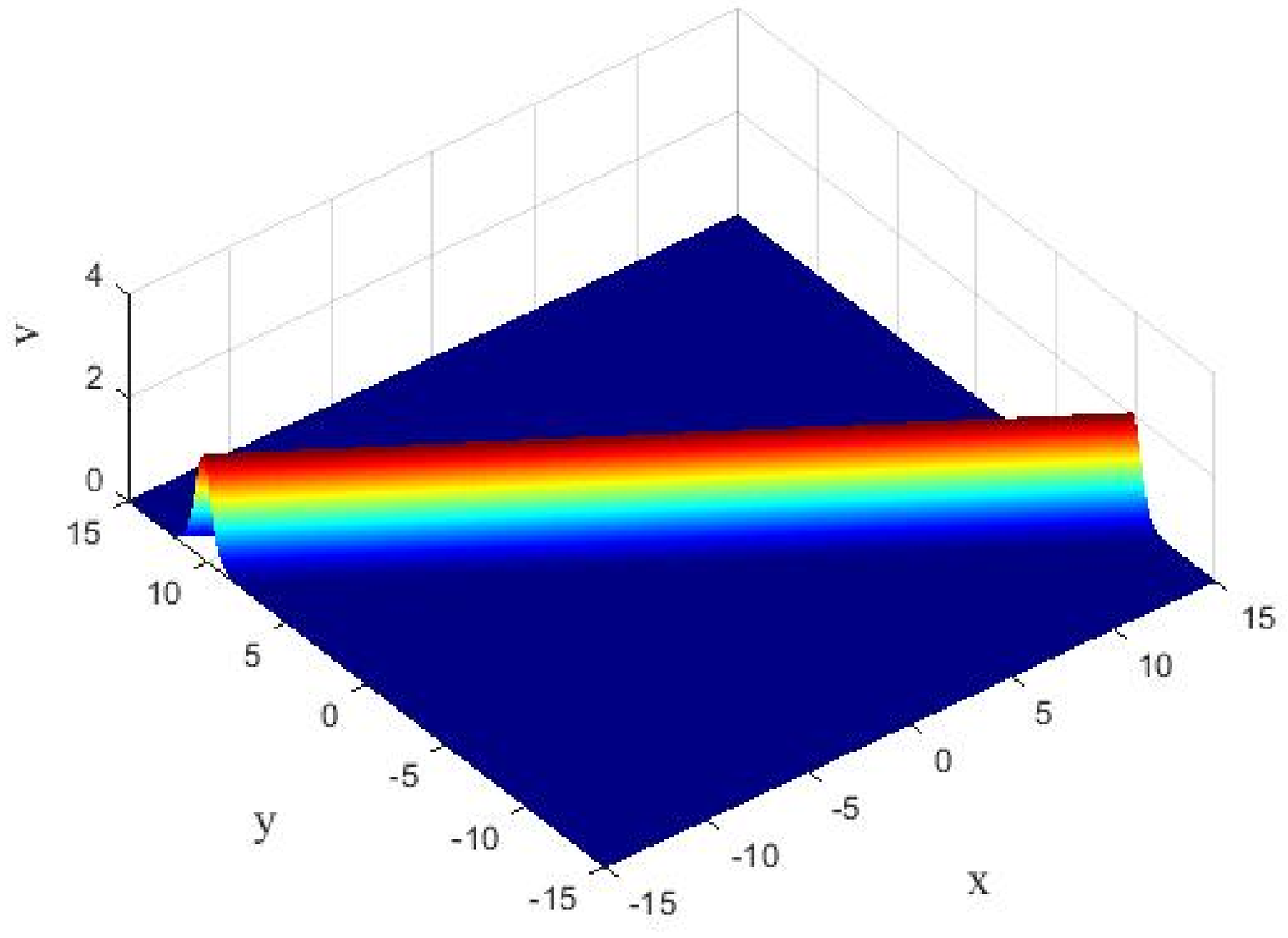}}
\centering
\caption{\small (color online) One dark-dark solitons in the degenerate case at the fixed time $t=0$ with the parameters chosen by  $k_1=1,k_2=2,m_1=1,n_1=1,\alpha_1=1,\alpha_2=1,\sigma_1=1,\sigma_2=-1,\beta_1=0,\beta_2=0,\xi_{10R}=0$.\label{xt8-f-1}  }
\end{figure}

(\romannumeral1) when $\alpha_1=\alpha_2$, then $H_1^{(1)}=H_1^{(2)}$ and $\psi_1^{(2)}=\psi_1^{(2)}$. In this case, the  two short wave components $\phi^{(1)}$ and $\phi^{(2)}$ have the same degrees of darkness and they are proportional to each other. At this moment, the dark-dark solitons of the coupled Maccari system including two short waves is same as  a scalar dark soliton in the Maccari system with one short wave. It can be seen as  a degenerate case similar to the coupled NLS \cite{8-16}, YO \cite{8-21}, Mel'nikov \cite{8-22} equations. The degenerate case for the dark-dark solitons is shown in Fig. \ref{xt8-f-1} , then we can find that the valleys for the two dark sloitons reach the zero plane and the long wave component possess bright  soliton.

\begin{figure}[H]
\renewcommand{\figurename}{{Fig.}}
\subfigure[]{\includegraphics[height=0.3\textwidth]{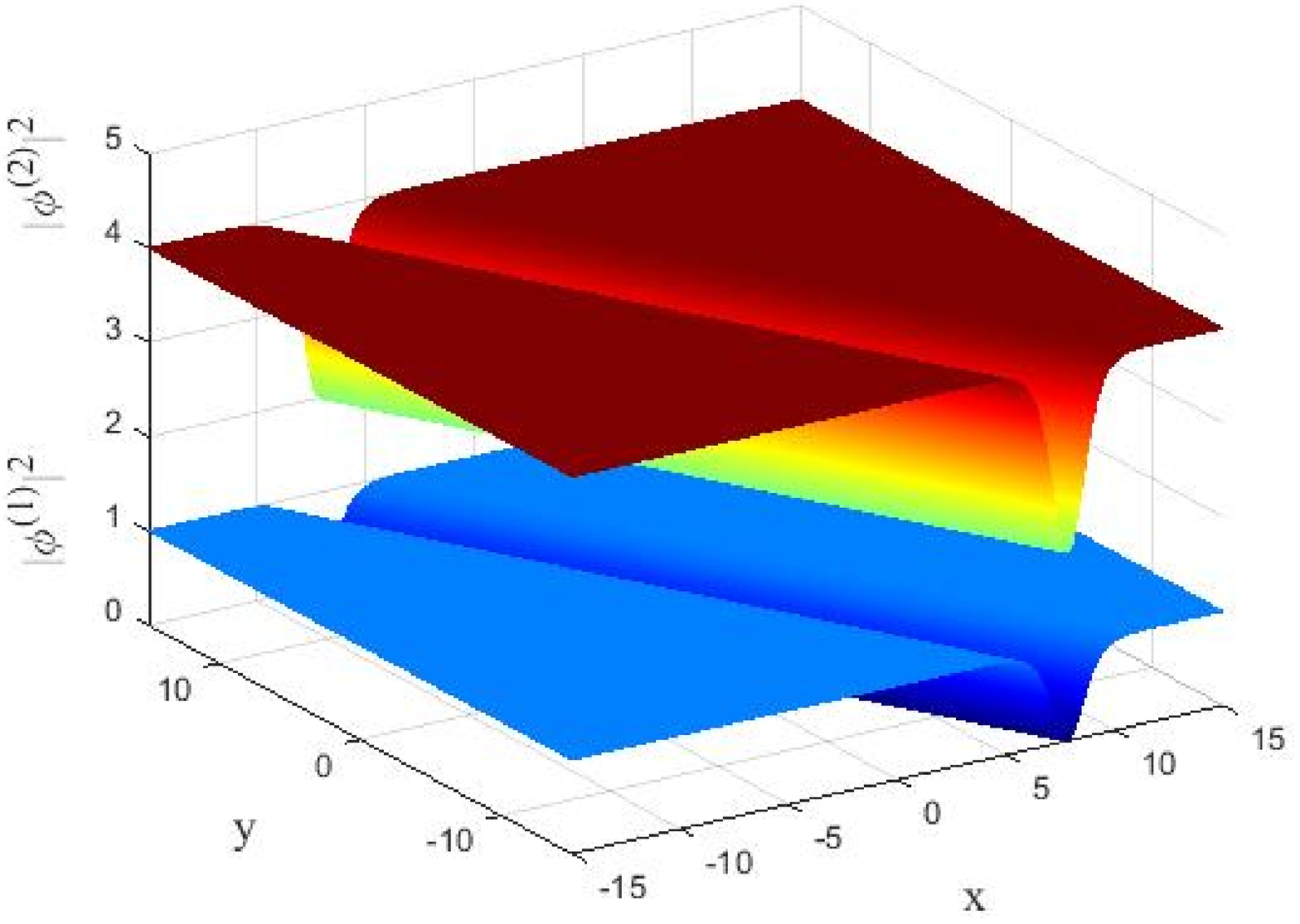}}
\centering
\subfigure[]{\includegraphics[height=0.3\textwidth]{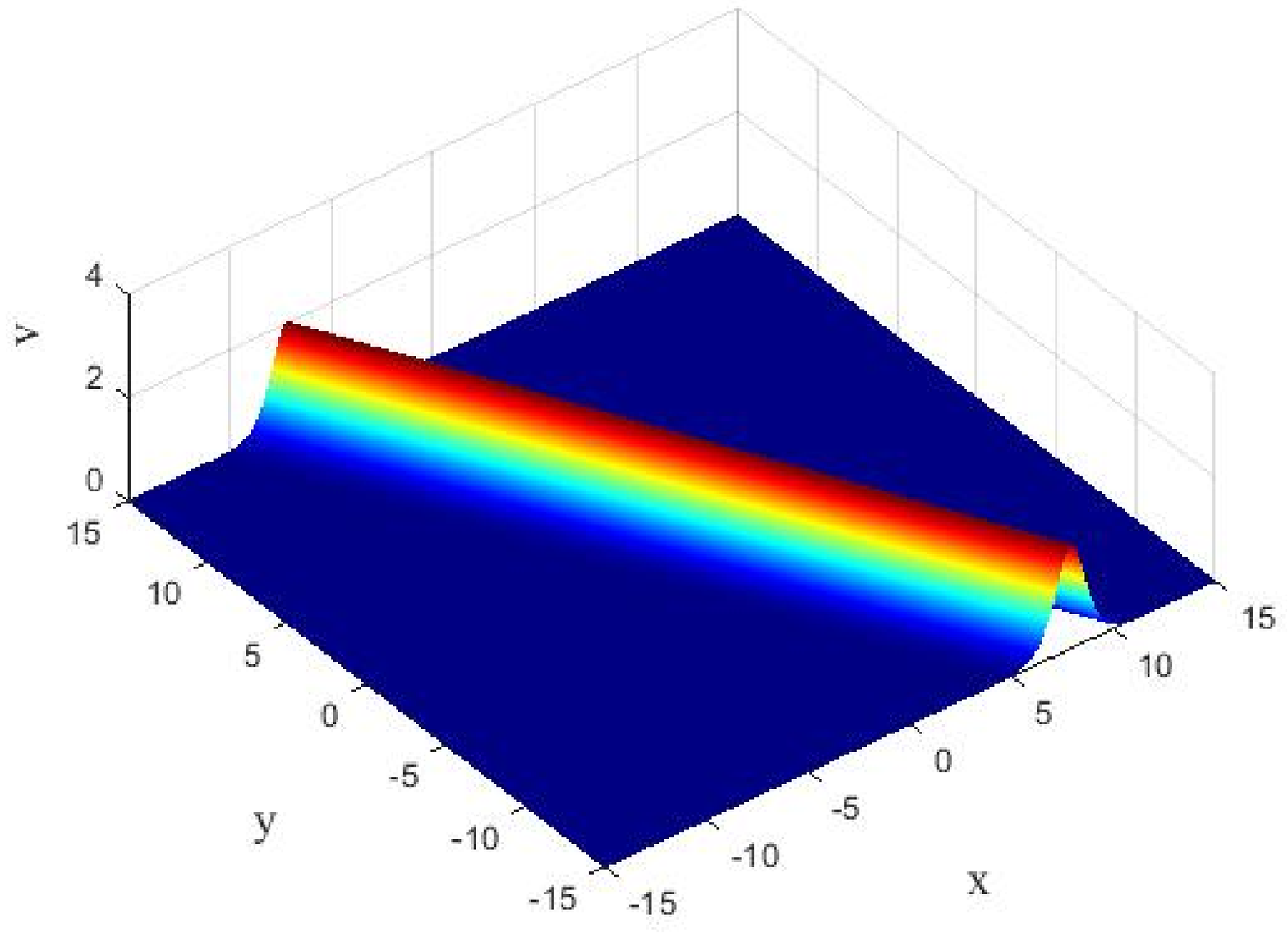}}
\centering
\caption{\small (color online) One dark-dark solitons in the non-degenerate case  at the fixed time $t=0$ with the parameters chosen by  $k_1=1,k_2=2,m_1=1,n_1=1,\alpha_1=1,\alpha_2=2,\sigma_1=1,\sigma_2=-1,\beta_1=0,\beta_2=0,\xi_{10R}=0$.\label{xt8-f-2}  }
\end{figure}

(\romannumeral2) when $\alpha_1\neq\alpha_2$, then $H_1^{(1)}\neq H_1^{(2)}$ and $\psi_1^{(2)}\neq\psi_1^{(2)}$. At this point, the long wave component $v$ still owns bright soliton and the  valley of the dark soliton in $\phi^{(1)}$ component reach the bottom, but the one in $\phi^{(2)}$ does not reach the bottom (which can be also called grey soliton), see Fig. \ref{xt8-f-2}. This is the non-degenerate case. From  Fig. \ref{xt8-f-2}(a), it is shown that the two short wave components  $\phi^{(1)}$ and $\phi^{(2)}$ have different degrees of darkness, and they are no longer proportional to each other. Furthermore, it is greatly different from the dark soliton in the Maccari system including one short wave components.

\subsection{Two dark-dark solitons}
Similarly, the two dark-dark solitons for the Maccari system (\ref{xt-8-1})-(\ref{xt-8-2}) can be directly calculated with $N=2$ in Eqs. (\ref{xt-8-17})-(\ref{xt-8-18}). The two dark-dark solitons can be written as follows
\begin{equation}
\phi^{(1)}=k_1e^{i(\alpha_1 x-\alpha_1^2+\beta_1(y))}\frac{h_2^{(1)}}{f_2},\quad
\phi^{(2)}=k_2e^{i(\alpha_2 x-\alpha_2^2+\beta_2(y))}\frac{h_2^{(2)}}{f_2},
\quad v=2(lnf_2)_{xx},\label{xt-8-24}
\end{equation}
where
\begin{eqnarray}
&& f_2=1+e^{\xi_1+\xi_1^{*}+\Theta_1}+e^{\xi_2+\xi_2^{*}+\Theta_2}+\Lambda_{12}
e^{\xi_1+\xi_1^{*}+\xi_2+\xi_2^{*}+\Theta_1+\Theta_2},\label{xt-8-23}\\
&&h_2^{(1)}=1+H_1^{(1)}e^{\xi_1+\xi_1^{*}+\Theta_1}+H_2^{(1)}e^{\xi_2+\xi_2^{*}+\Theta_2}+H_1^{(1)}H_2^{(1)}\Lambda_{12}
e^{\xi_1+\xi_1^{*}+\xi_2+\xi_2^{*}+\Theta_1+\Theta_2},\\
&&h_2^{(2)}=1+H_1^{(2)}e^{\xi_1+\xi_1^{*}+\Theta_1}+H_2^{(2)}e^{\xi_2+\xi_2^{*}+\Theta_2}+H_1^{(2)}H_2^{(2)}\Lambda_{12}
e^{\xi_1+\xi_1^{*}+\xi_2+\xi_2^{*}+\Theta_1+\Theta_2},\label{xt-8-25}
\end{eqnarray}
with
\begin{eqnarray}
&&e^{\Theta_j}=\frac{1}{p_j+p_j^{*}}=\frac{1}{2m_j},\quad H_j^{(1)}=-\frac{p_j-i\alpha_1}{p_j^{*}+i\alpha_1}=-\frac{m_j+i(n_j-\alpha_1)}{m_j-i(n_j-\alpha_1)},\\
&&H_j^{(2)}{=}{-}\frac{p_j-i\alpha_2}{p_j^{*}+i\alpha_2}{=}{-}\frac{m_j+i(n_j-\alpha_2)}{m_j-i(n_j-\alpha_2)},\quad
\Lambda_{12}{=}\left| \frac{p_1-p_2}{p_1+p_2^{*}}\right|^2{=}\frac{(m_1-m_2)^2+(n_1-n_2)^2}{(m_1+m_2)^2+(n_1-n_2)^2},\\
&&\nonumber\xi_j+\xi_j^{*}=2m_j x-4m_j n_j t-(\frac{\sigma_1k_1^2m_j}
{m_j^2+(n_j-\alpha_1)^2}+\frac{\sigma_2k_2^2m_j}{m_j^2+(n_j-\alpha_2)^2})y+2\xi_{j0R}\\
&&\hspace{1.1cm}\equiv L _{jx}x+L_{jt}t+L_{jy}y~(j=1,2),\label{xt-8-27}
\end{eqnarray}
here, $p_j=m_j+in_j$,~$\xi_{j0}=\xi_{j0R}+i\xi_{j0I}$ and $m_j,n_j,\alpha_1,\alpha_2,\xi_{j0R},\xi_{j0I}$ are real constants.

From the above expression of $\Lambda_{12}$, we find that the denominator becomes zero with the parameters chosen by $m_1=-m_2,n_1=n_2$. At this point, the two solitons interaction exist Y-shape, which is called the resonant solitons \cite{8-21,8-22}.  When $\Lambda_{12}$ exists, we discuss the asymptotic analysis for the two silitons as follows.

(1) Before the interaction ($x,y\rightarrow -\infty$):

Soliton 1 $s_1$ ($\xi_{1R}\simeq0,\xi_{2R}\rightarrow -\infty$)
\begin{eqnarray*}
&&\phi_1^{(1)-}\simeq\frac{k_1}{2}e^{i\theta_1}[(1+H_1^{(1)}+(H_1^{(1)}-1))tanh(\frac{\xi_1+\xi_1^{*}+\Theta_1}{2})],\\
&&\phi_1^{(2)-}\simeq\frac{k_2}{2}e^{i\theta_2}[(1+H_1^{(2)}+(H_1^{(2)}-1))tanh(\frac{\xi_1+\xi_1^{*}+\Theta_1}{2})],\\
&&v_1^{-}\simeq\frac{(p_1+p_1^{*})^2}{2}sech^{2}(\frac{\xi_1+\xi_1^{*}+\Theta_1}{2}),
\end{eqnarray*}

Soliton 2 $s_2$ ($\xi_{2R}\simeq0,\xi_{1R}\rightarrow +\infty$)
\begin{eqnarray*}
&&\phi_2^{(1)-}\simeq\frac{k_1H_1^{(1)}}{2}e^{i\theta_1}[(1+H_2^{(1)})+(H_2^{(1)}-1)tanh(\frac{\xi_2+\xi_2^{*}+
\Theta_2+ln\Lambda_{12}}{2})],\\
&&\phi_2^{(2)-}\simeq \frac{k_2H_1^{(2)}}{2}e^{i\theta_2}[(1+H_2^{(2)})+(H_2^{(2)}-1)tanh(\frac{\xi_2+\xi_2^{*}+
\Theta_2+ln\Lambda_{12}}{2})],\\
&&v_2^{-}\simeq\frac{(p_2+p_2^{*})^2}{2}sech^{2}(\frac{\xi_2+\xi_2^{*}+\Theta_2+ln\Lambda_{12}}{2}),
\end{eqnarray*}

(2) After the interaction ($x,y\rightarrow +\infty$):

Soliton 1 $s_1$ ($\xi_{1R}\simeq0,\xi_{2R}\rightarrow +\infty$)
\begin{eqnarray*}
&&\phi_1^{(1)+}\simeq\frac{k_1H_2^{(1)}}{2}e^{i\theta_1}[(1+H_1^{(1)})
+(H_1^{(1)}-1)tanh(\frac{\xi_1+\xi_1^{*}+\Theta_1+ln\Lambda_{12}}{2})],\\
&&\phi_1^{(2)+}\simeq\frac{k_2H_2^{(2)}}{2}e^{i\theta_2}[(1+H_1^{(2)})
+(H_1^{(2)}-1)tanh(\frac{\xi_1+\xi_1^{*}+\Theta_1+ln\Lambda_{12}}{2})],\\
&&v_1^{+}\simeq\frac{(p_1+p_1^{*})^2}{2}sech^{2}(\frac{\xi_1+\xi_1^{*}+\Theta_{1}+ln\Lambda_{12}}{2}),
\end{eqnarray*}

Soliton 2 $s_2$ ($\xi_{2R}\simeq0,\xi_{1R}\rightarrow -\infty$)
\begin{eqnarray*}
&&\phi_2^{(1)+}\simeq\frac{k_1}{2}e^{i\theta_1}[(1+H_2^{(1)})+(H_2^{(1)}-1)tanh(\frac{\xi_2+\xi_2^{*}+\Theta_2}{2})],\\
&&\phi_2^{(2)+}\simeq \frac{k_2}{2}e^{i\theta_2}[(1+H_2^{(2)})+(H_2^{(2)}-1)tanh(\frac{\xi_2+\xi_2^{*}+\Theta_2}{2})],\\
&&v_2^{+}\simeq\frac{(p_2+p_2^{*})^2}{2}sech^{2}(\frac{\xi_2+\xi_2^{*}+\Theta_2}{2}),
\end{eqnarray*}
where $\xi_j=\xi_{jR}+i\xi_{jI}~(j=1,2)$, $\phi_j^{(1)-},\phi_j^{(2)-}$ denotes the two solitons for the two short wave components before collision, and $v_j^{-}$ are the two solitons for long wave component before collision. Conversely, $\phi_j^{(1)+},\phi_j^{(2)+}$ and $v_j^{+}$ stand for the two solitons in the corresponding components after collision.

From the above expressions of asymptotic analysis, we can directly calculate that
\begin{eqnarray}
&&  A_1^{(1){-}}{=}A_1^{(1){+}}{=}|k_1|(1{-}\frac{\sqrt{\tfrac{1}{3}m_1^2{+}(n_1{-}\alpha_1)^2}}{m_1^2{+}(n_1{-}\alpha_1)^2}),  A_1^{(2){-}}{=}A_1^{(2){+}}{=}|k_2|(1{-}\frac{\sqrt{\tfrac{1}{3}m_1^2{+}(n_1{-}\alpha_1)^2}}{m_1^2{+}(n_1{-}\alpha_1)^2}),\label{xt-8-21}\\
&&A_2^{(1)-}{=}A_2^{(1)+}{=}|k_1|(1{-}\frac{\sqrt{\tfrac{1}{3}m_2^2{+}(n_2{-}\alpha_1)^2}}{m_2^2{+}(n_2{-}\alpha_1)^2}),
A_2^{(2)-}{=}A_2^{(2){+}}{=}|k_2|(1{-}\frac{\sqrt{\tfrac{1}{3}m_2^2{+}(n_2{-}\alpha_2)^2}}{m_2^2{+}(n_2{-}\alpha_2)^2}),\\
&&A_v^{(1)-}{=}A_v^{(1)+}{=}\frac{(p_1+p_1^*)^2}{2}{=}2m_1^2,\quad A_v^{(2)-}{=}A_v^{(2)+}{=}\frac{(p_2+p_2^*)^2}{2}{=}2m_2^2,\label{xt-8-22}
\end{eqnarray}
Where $A_1^{(j)-}$,~$A_1^{(j)+}~(j=1,2)$ denote the amplitude of the soliton $j$ ($s_j$) in the short wave component $\phi^{(1)}$ before and after collision, respectively. Analogously, $A_2^{(j)-}$,~$A_2^{(j)+}$ are the amplitudes of the soliton $j$ in the short wave component $\phi^{(2)}$ before and after collision. Additionally, the amplitudes of the corresponding soliton $s_j$ in the long wave component $v$ can be written as $A_v^{(j)-}$ (before collision) and $A_v^{(j)+}$ (after collision). From the asymptotic analysis and Eqs. (\ref{xt-8-21})-(\ref{xt-8-22}), we can make a conclusion that the amplitude and velocity of each soliton maintain unchange during the interaction except for some phase shifts, and the collisions in the three components are all elastic.

\begin{figure}[!ht]
\renewcommand{\figurename}{{Fig.}}
\subfigure[]{\includegraphics[height=0.24\textwidth]{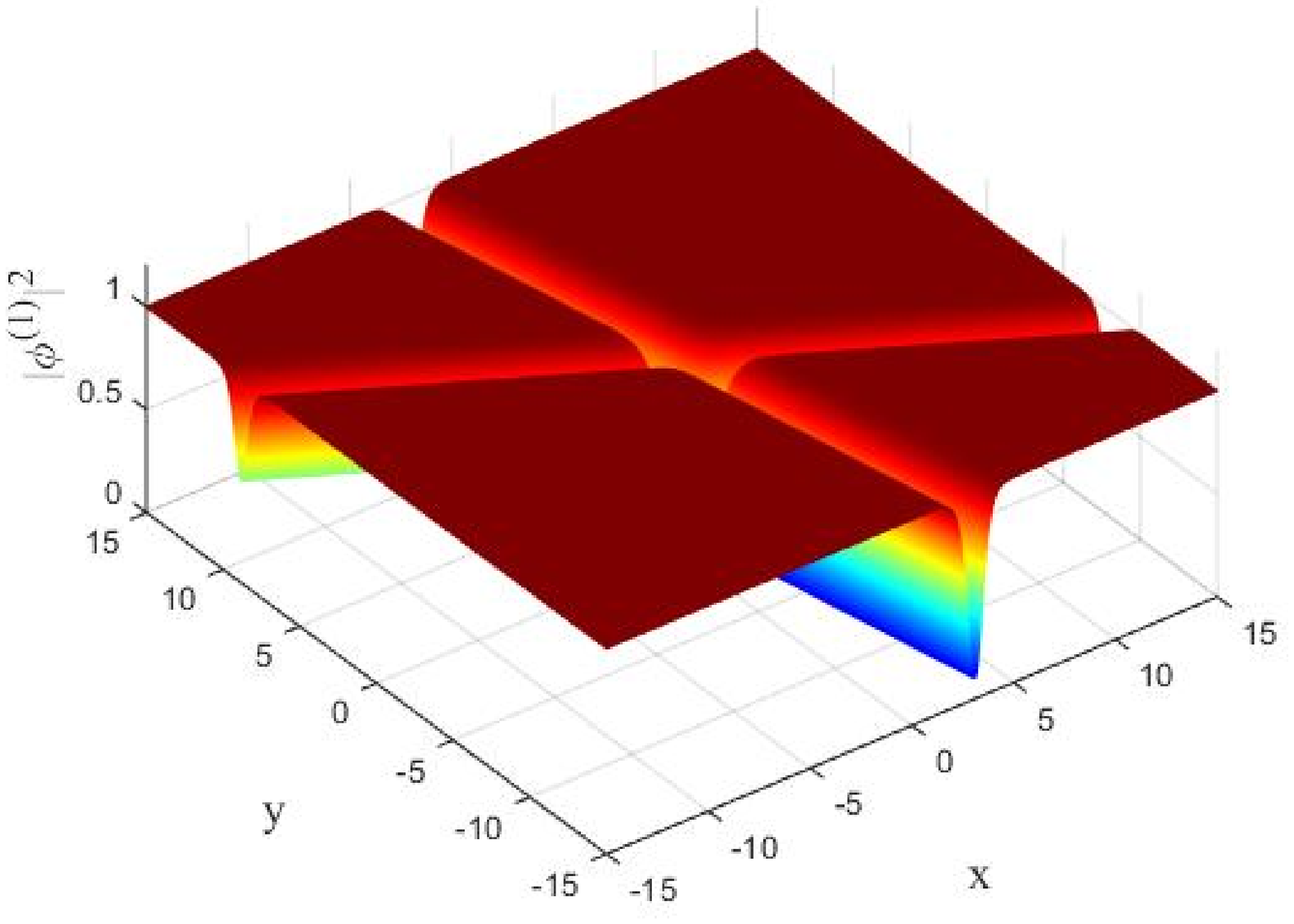}}
\centering
\subfigure[]{\includegraphics[height=0.24\textwidth]{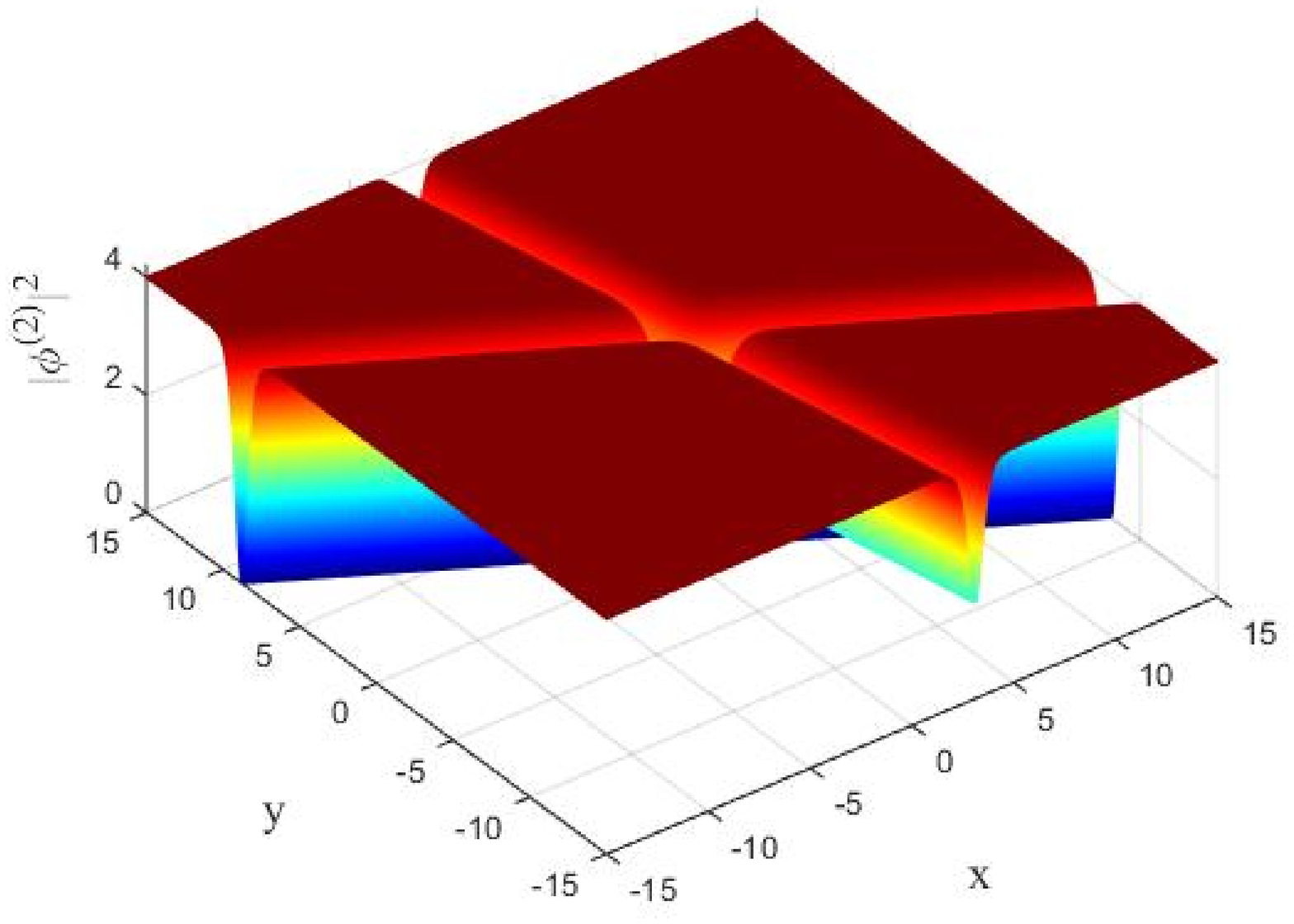}}
\centering
\subfigure[]{\includegraphics[height=0.24\textwidth]{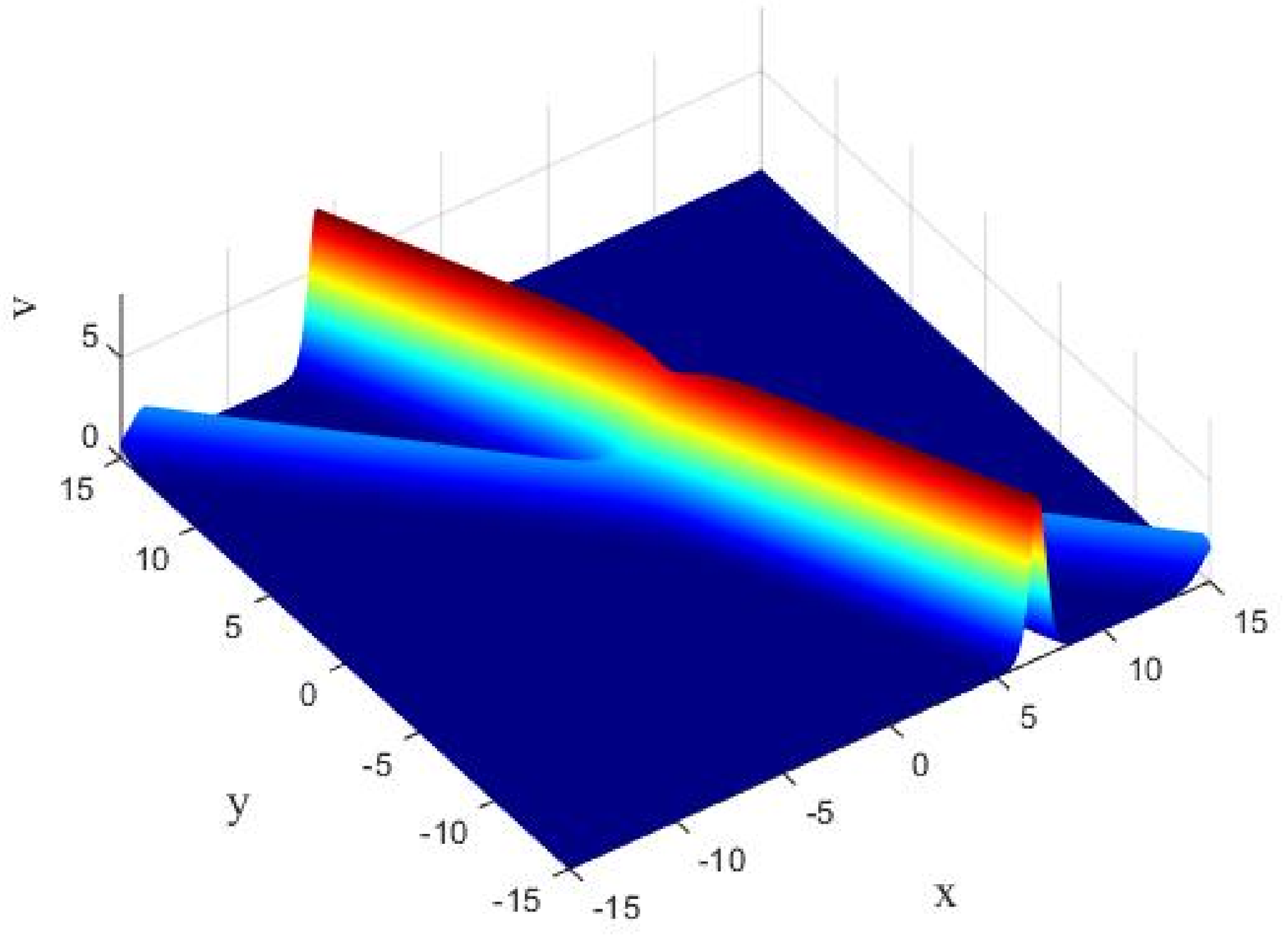}}
\caption{\small (color online) Two dark-dark solitons  at the fixed time $t=0$ with the parameters chosen by  $k_1=1,k_2=2,m_1=1,n_1=2,m_2=2,n_2=\frac{1}{3},\alpha_1=1,\alpha_2=2,\sigma_1=1,\sigma_2=-1,\beta_1=0,\beta_2=0,\xi_{10R}=0,\xi_{20R}=0$.\label{xt8-f-3}  }
\end{figure}

From Eq. (\ref{xt-8-23}), we can easily get a series of  nonsingular two dark-dark soliton solutions with the parameters chosen by $m_j>0~(j=1,2)$.  Meantime, the collision of two dark-dark solitons is exhibited in Fig. \ref{xt8-f-3}. It is shown that the two solitons interact with each other without any change of amplitude, darkness and velocity in the three components after collision. There is no energy exchange between the two short wave components or two dark solitons in the same component after interaction. This kind of transmission of energy of dark-dark soliton solutions occurs in all possible cases of  nonlinearity coefficients $\sigma_1$ and $\sigma_2$ (all positive, all negative and mixed).

For the two bright solitons in the long wave component $v$, we can discuss them in two types (O-type and P-type) according to asymptotic analysis. From Eq. (\ref{xt-8-22}), we can find that the amplitudes of the two solitons $s_1$ and $s_2$ for the long wave component in the asymptotic analysis are $2m_1^2$ and $2m_2^2$, respectively. According the classifications of the soliton solutions in the Kadomtsev-Petviashvili (KP) equation \cite{8-41,8-42}, the two bright solitons in $v$ can be classified as O-type (`O' for original) and P-type (`P' for physical) soliton interactions.

(\romannumeral1) When $m_1m_2<0$, then $\Lambda_{12}>1$, this type is called O-type soliton interaction. The two asymptotic soliton amplitudes $2m_1^2$ and $2m_2^2$ for the long wave component $v$ can be equivalent when $m_1=-m_2$. It is found that the maximum of $v$ (interaction peak) is always bigger than the sum of the asymptotic soliton ($s_1$ and $s_2$) amplitudes.

(\romannumeral2) When $m_1m_2>0$, then $0<\Lambda_{12}<1$, this type is called P-type soliton interaction. The following equality  that $2m_1^2=2m_2^2~(m_1\neq m_2)$ is not possible in this case. It is  shown that the maximum of $v$ (interaction peak) is always  smaller than the sum of the asymptotic soliton amplitudes. The two bright solitons in Fig.\ref{xt8-f-3}(c) is just P-type soliton interaction.

 It should be  noted that the two different types of soliton interaction do not depend on the imaginary parts of $p_1$ and $p_2$ (namely $n_1$ and $n_2$). The  above mentioned Y-shape soliton solution can be constructed through using a limiting process $n_1\rightarrow n_2$  in the O-type two solitons. Similar with the cases in one- and two-dark-dark solitons, the higher-order dark-dark solitons for the Maccari system (\ref{xt-8-1})-(\ref{xt-8-2}) can also be discussed and we omit them here.

\section{Soliton bound states}
In the following contents, we will consider the soliton bound states in the three components $\phi^{(1)}$,~$\phi^{(2)}$ and $v$. In order to construct the bound states for the Maccari system (\ref{xt-8-1})-(\ref{xt-8-2}), the solitons of the short and long wave components should possess the same velocities along the $x$- and $y$-direction. From the concrete expressions (\ref{xt-8-24})-(\ref{xt-8-25}) of two-dark-dark solitons, we can directly calculate the two velocities $V_{jx}$~(along the $x$-direction) and $V_{jy}$~(along the $y$-direction) as follows ($j=1,2$ denote the two different solitons)
\begin{eqnarray}
V_{jx}=2n_j,\quad V_{j,y}=-\frac{4n_j}{\frac{\sigma_1k_1^2}{m_j^2+(n_j-\alpha_1)^2}+\frac{\sigma_2k_2^2}{m_j^2+(n_j-\alpha_2)^2}}~(j=1,2).\label{xt-8-26}
\end{eqnarray}

 When $V_{jx}=V_{jy}=0$, we can get the stationary dark-dark soliton bound states. From Eq. (\ref{xt-8-26}), it only requires that $n_1=n_2=0$. There are not other constrains for the nonlinear coefficients, then all values of $\sigma_1$,$\sigma_2$ (all positive, all negative and mixed) can satisfy this condition. However, the relative constrains only exist when the nonlinearity coefficients take opposite signs in the coupled NLS equations \cite{8-16} and YO equations \cite{8-21}. The higher-order
stationary dark-dark soliton bound states only require that $n_1=n_2=\cdots=n_i~(3\leq i\leq N)$, and we can make a conclusion that the bound states can exist up to arbitrary order in the stationary case.

Accoring to the relationship between the coefficients $L_{jx}$ and $L_{jy}$ in Eq. (\ref{xt-8-27}), the stationary bound states can be classified as two types---oblique and parallel.

\begin{figure}[H]
\renewcommand{\figurename}{{Fig.}}
\subfigure[]{\includegraphics[height=0.24\textwidth]{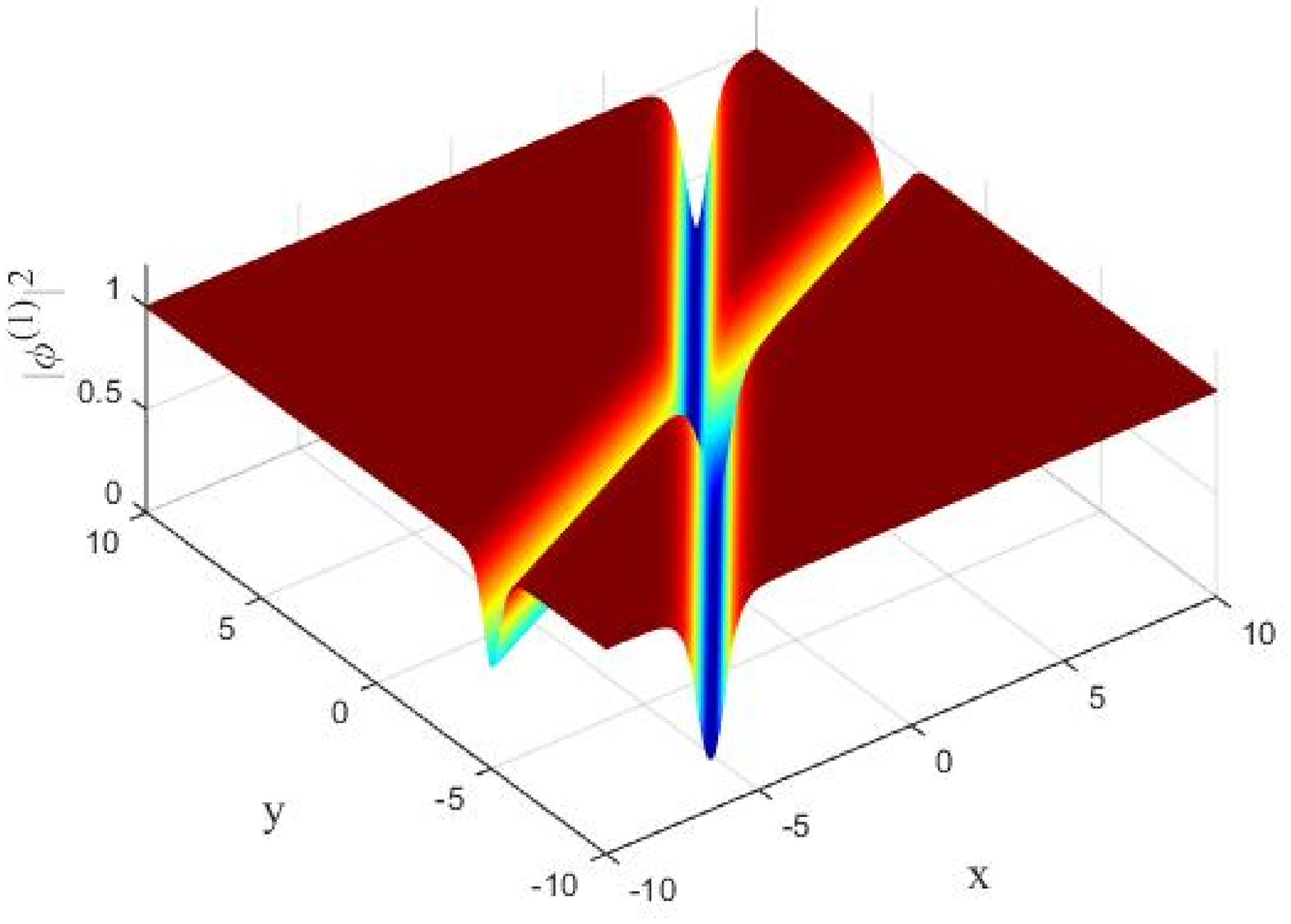}}
\centering
\subfigure[]{\includegraphics[height=0.24\textwidth]{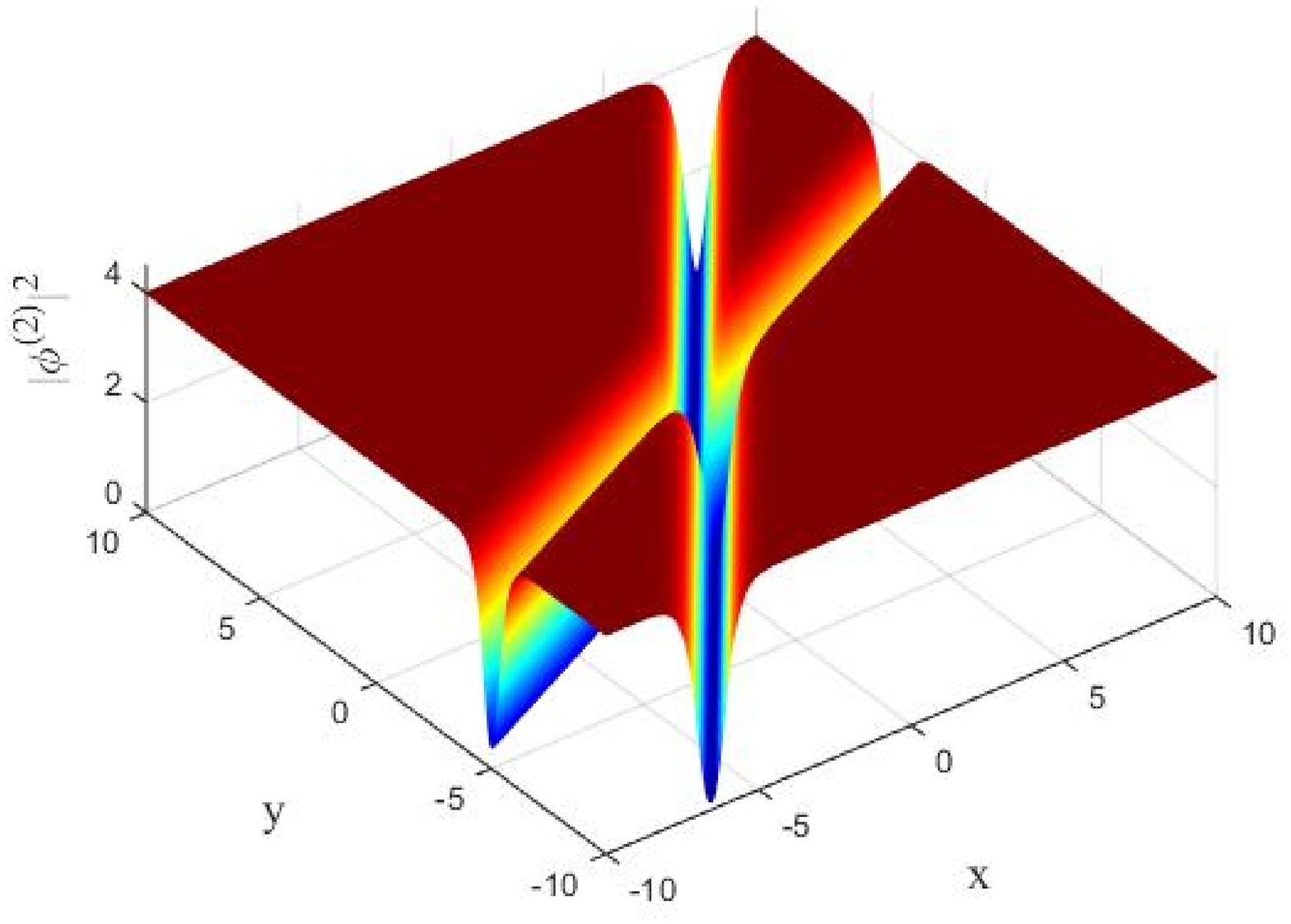}}
\centering
\subfigure[]{\includegraphics[height=0.24\textwidth]{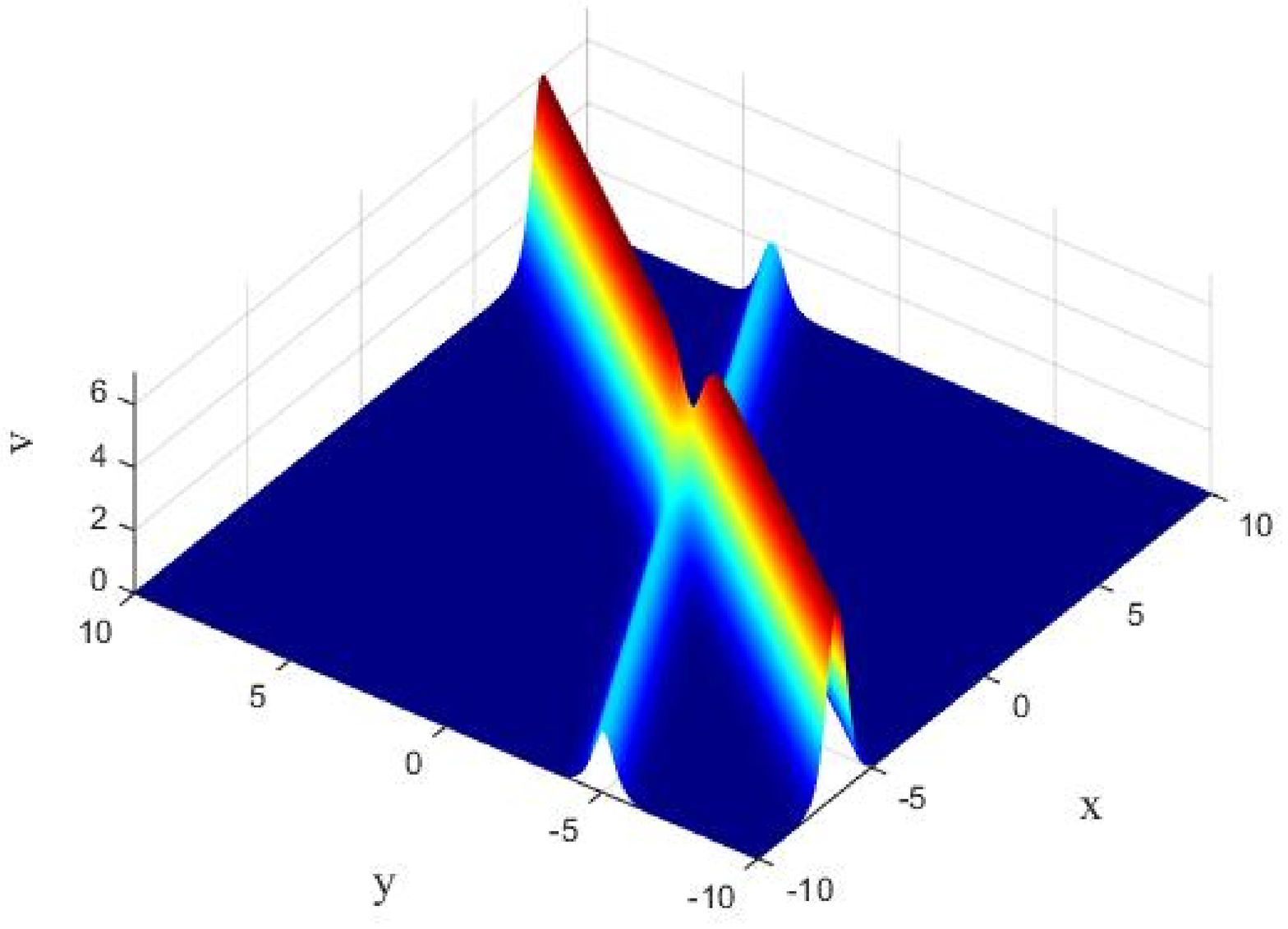}}
\caption{\small (color online) The stationary two dark-dark soliton bound states in the oblique case with the parameters chosen by  $k_1=1,k_2=2,m_1=1,n_1=0,m_2=\sqrt{3},n_2=0,\alpha_1=1,\alpha_2=\frac{1}{3},\sigma_1=1,\sigma_2=1,\beta_1=0,\beta_2=0,\xi_{10R}=0,\xi_{20R}=0$.\label{xt8-f-4}  }
\end{figure}

(a) If $\frac{L_{1y}}{L_{1x}}\neq\frac{L_{2y}}{L_{2x}}$, namely $\frac{\sigma_1k_1^2}{m_1^2+\alpha_1^2}+
\frac{\sigma_2k_2^2}{m_1^2+\alpha_2^2}\neq\frac{\sigma_1k_1^2}{m_2^2+\alpha_1^2}+
\frac{\sigma_2k_2^2}{m_2^2+\alpha_2^2}$, this is the oblique stationary bound state, see Fig.\ref{xt8-f-4}. Here, it is shown that the coefficients of the independent variable $t$ are all zero in the expressions of two-dark-dark solitons, and these two solitons do not propagate along time $t$. Moreover, the oblique stationary bound state can be generated for all possible combinations of nonlinearity coefficients $\sigma_1,\sigma_2$ consisted of positive, negative and mixed cases.

\begin{figure}[H]
\renewcommand{\figurename}{{Fig.}}
\subfigure[]{\includegraphics[height=0.24\textwidth]{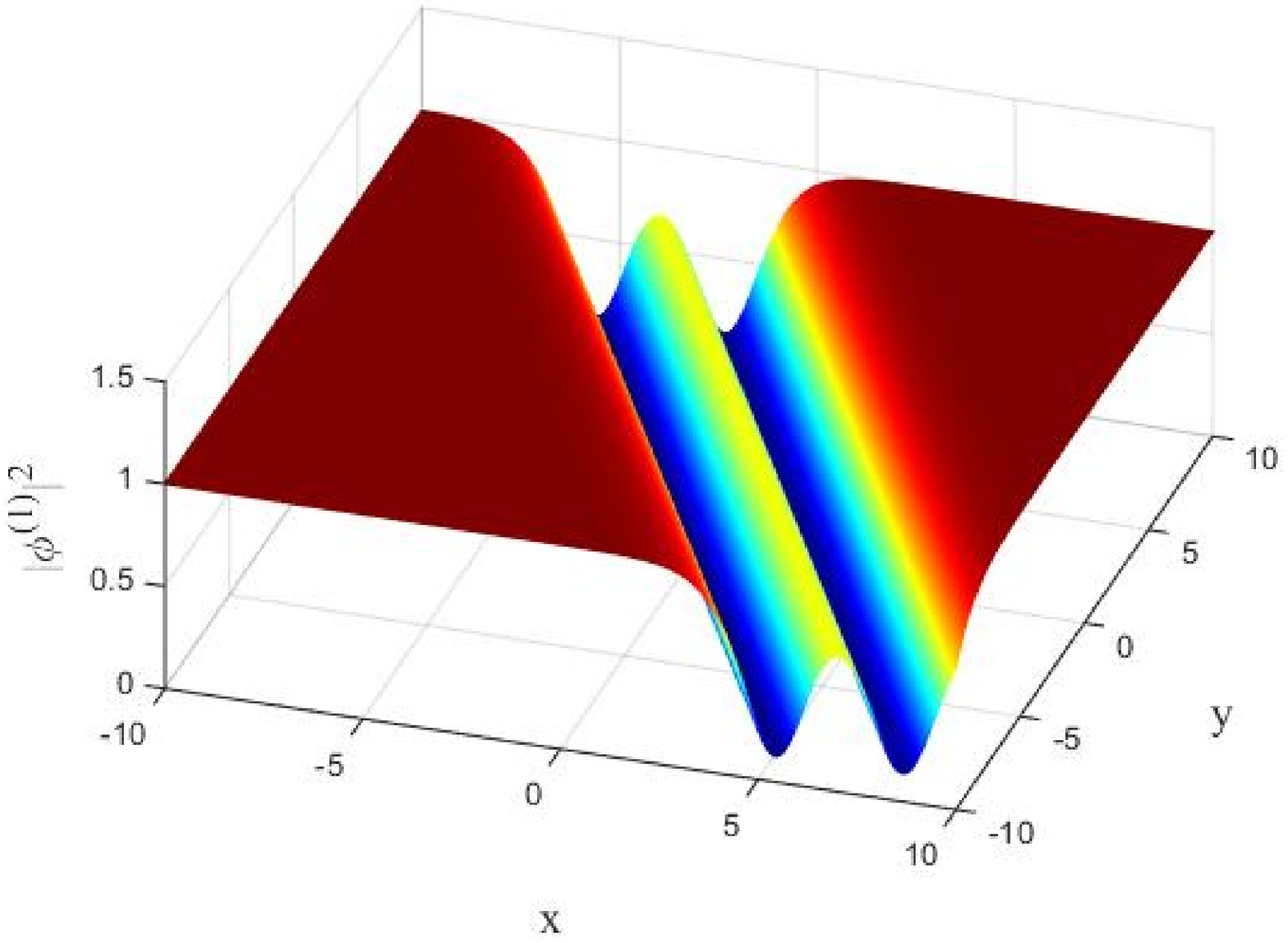}}
\centering
\subfigure[]{\includegraphics[height=0.24\textwidth]{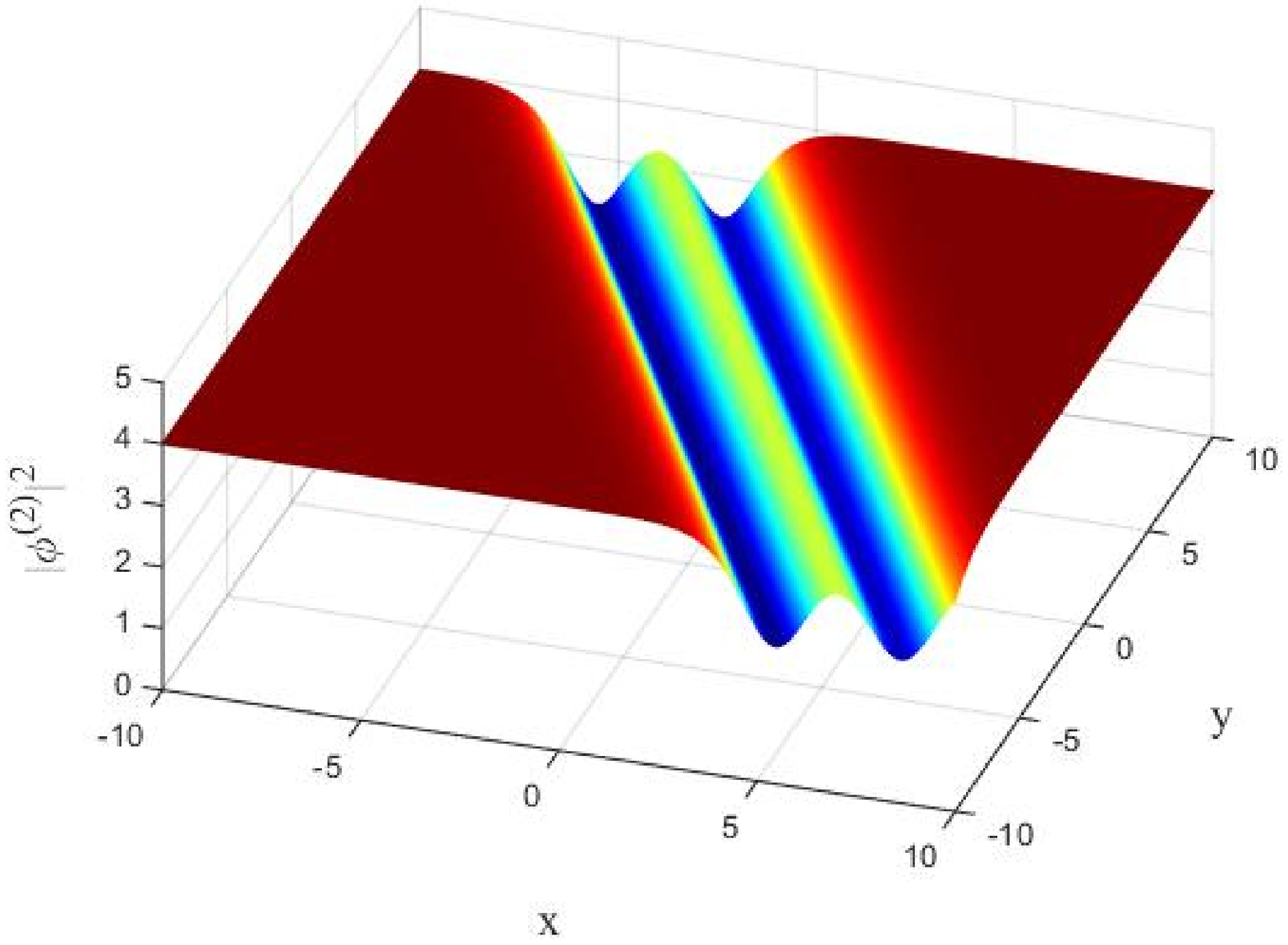}}
\centering
\subfigure[]{\includegraphics[height=0.24\textwidth]{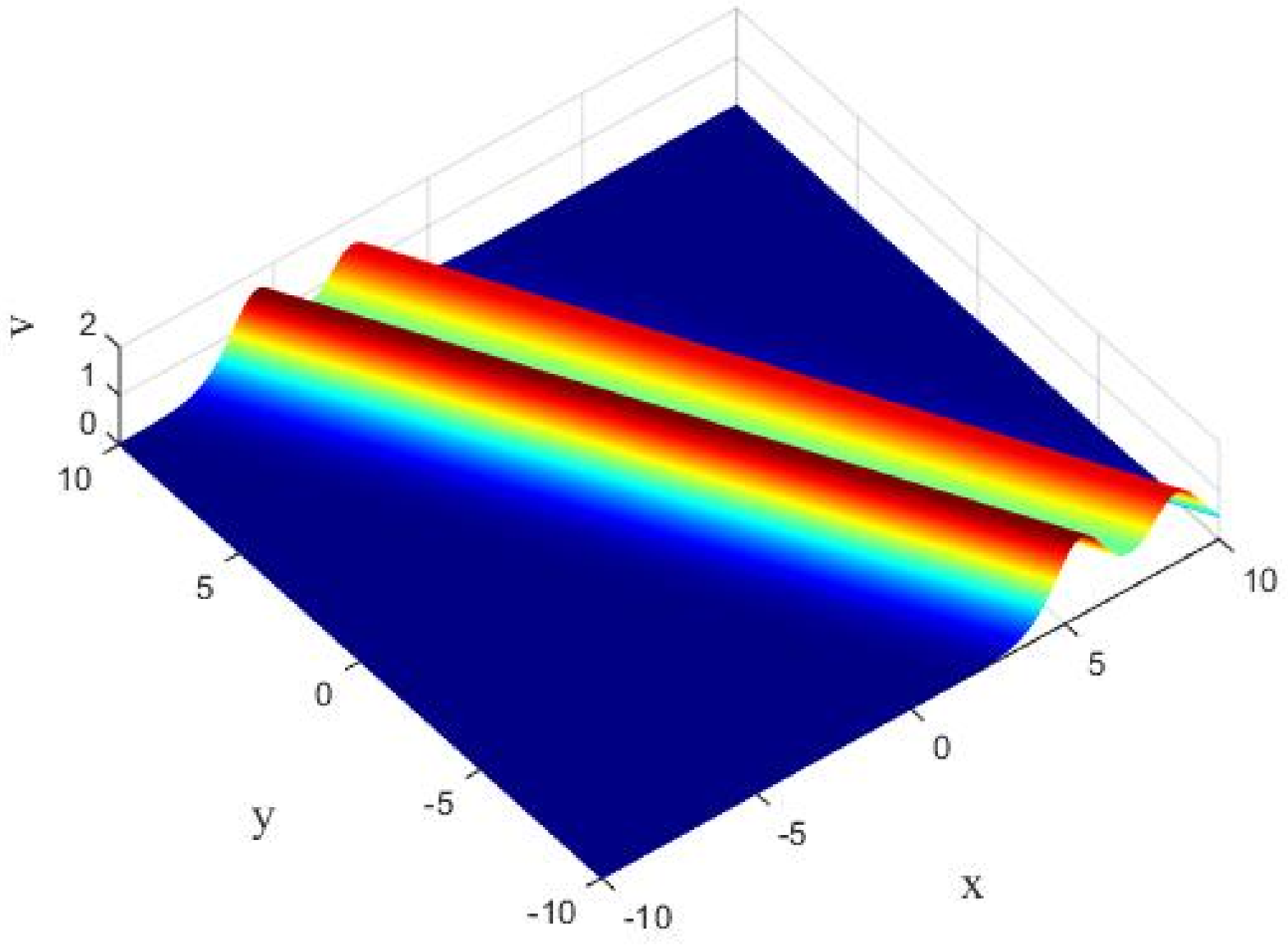}}
\caption{\small (color online) The stationary two dark-dark soliton bound states in the parallel case with the parameters chosen by  $k_1=1,k_2=2,m_1=1,n_1=0,m_2=\frac{1}{33}\sqrt{671},n_2=0,\alpha_1=\frac{1}{3},\alpha_2=1,\sigma_1=1,\sigma_2=-1,\beta_1=0,\beta_2=0,\xi_{10R}=0,\xi_{20R}=0$.\label{xt8-f-5}  }
\end{figure}

(b)If $\frac{L_{1y}}{L_{1x}}=\frac{L_{2y}}{L_{2x}}$, namely $\frac{\sigma_1k_1^2}{m_1^2+\alpha_1^2}+
\frac{\sigma_2k_2^2}{m_1^2+\alpha_2^2}=\frac{\sigma_1k_1^2}{m_2^2+\alpha_1^2}+
\frac{\sigma_2k_2^2}{m_2^2+\alpha_2^2}$, this is the parallel stationary bound state, see Fig. \ref{xt8-f-5}. From Eq. (\ref{xt-8-23}), one of the constraint conditions for constructing the nonsingular two-dark-dark solitons is that $m_1>0,m_2>0$. Given the following function $f(x)=\frac{\sigma_1k_1^2}{x^2+\alpha_1^2}+\frac{\sigma_1k_1^2}{x^2+\alpha_2^2}~(x>0)$, and the  parallel stationary bound state exists if and only if  $f(m_1)=f(m_2)$ by choosing appropriate values of $\sigma_1,\sigma_2,k_1,k_2,\alpha_1,\alpha_2$. Besides, the function $f(x)$ is  monotonous  in  the interval $x>0$ when the nonlinearity coefficients $\sigma_1,\sigma_2$ take the same signs (all positive or all negative). Considering the above condition $m_1>0,m_2>0$, it is not possible that $f(m_1)=f(m_2)$ when $m_1\neq m_2,\sigma_1\sigma_2>0$. Based on the above facts, we find that the parallel stationary is only possible when nonlinearity coefficients $\sigma_1,\sigma_2$ take opposite signs.

When $V_{jx}=V_{jy}\neq0$, the moving two dark-dark soliton bound states can be constructed. From Eq. (\ref{xt-8-26}), the following restricted conditions for constructing moving bound states can be written as
\begin{equation}
V_{jx}: n_1{=}n_2\neq 0,\quad V_{jy}: \frac{\sigma_1k_1^2}{m_1^2+(n_1-\alpha_1)^2}{+}\frac{\sigma_2k_2^2}{m_1^2+(n_1-\alpha_2)^2}{=}
\frac{\sigma_1k_1^2}{m_2^2+(n_1-\alpha_1)^2}{+}\frac{\sigma_2k_2^2}{m_2^2+(n_1-\alpha_2)^2}.
\end{equation}

Similar with the case in parallel stationary bound state, we can also construct the function $g(x)=\frac{\sigma_1k_1^2}{x^2+(n_1-\alpha_1)^2}+\frac{\sigma_2k_2^2}{x^2+(n_1-\alpha_2)^2}~(x>0)$. The function $g(x)$ is monotonous in the interval $x>0$ with fixed parameters $k_1,k_2,n_1,\alpha_1,\alpha_2$ when nonlinearity coefficients $\sigma_1,\sigma_2$ take the same signs. Considering the nonsingular condition $m_1>0,m_2>0$, it is not possible that $g(m_1)=g(m_2)~(m_1\neq m_2)$ if $\sigma_2\sigma_2>0$. It is shown that the moving bound states are only possible when nonlinearity coefficients $\sigma_1,\sigma_2$ take opposite signs.

\begin{figure}[H]
\renewcommand{\figurename}{{Fig.}}
\subfigure[]{\includegraphics[height=0.24\textwidth]{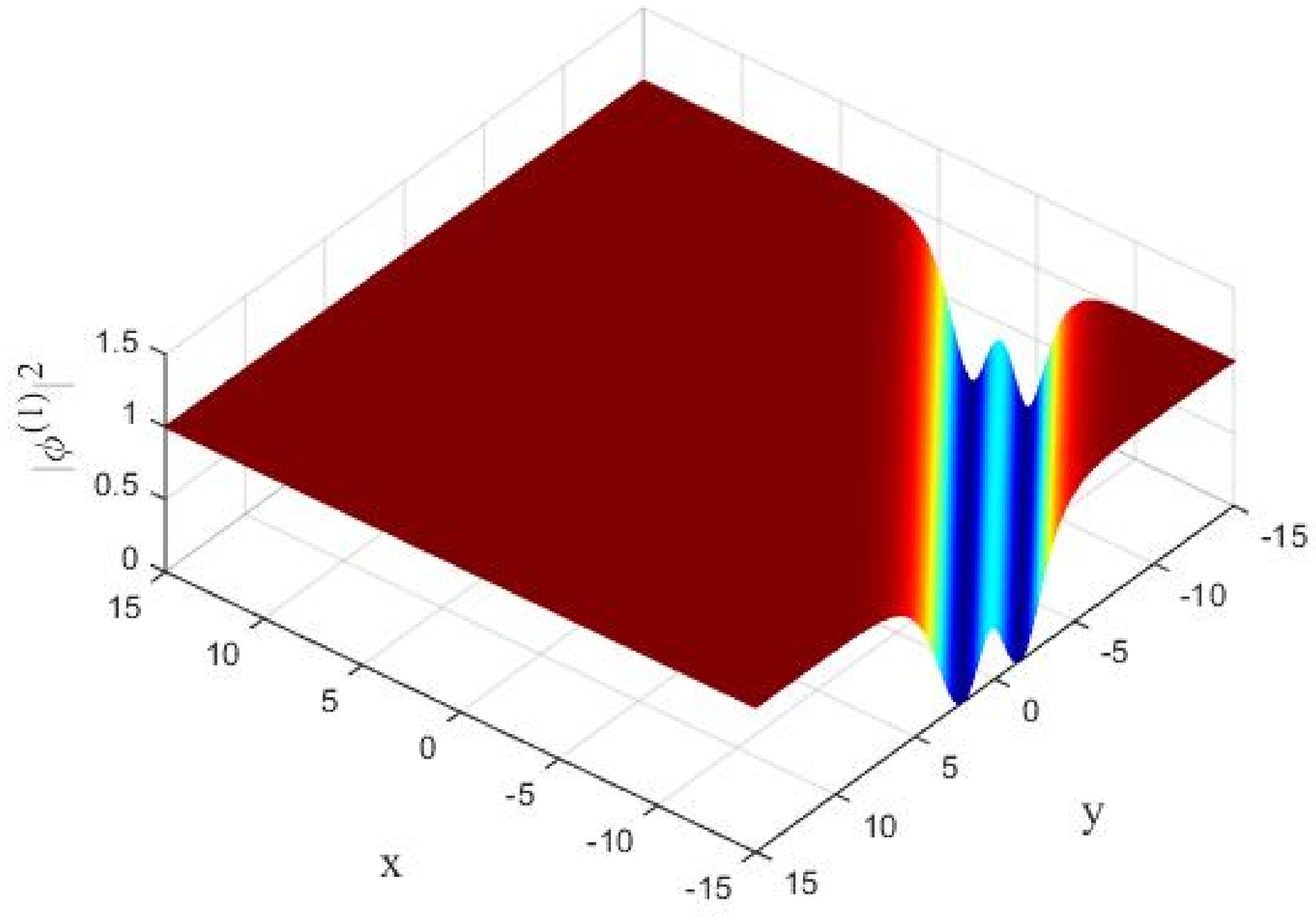}}
\centering
\subfigure[]{\includegraphics[height=0.24\textwidth]{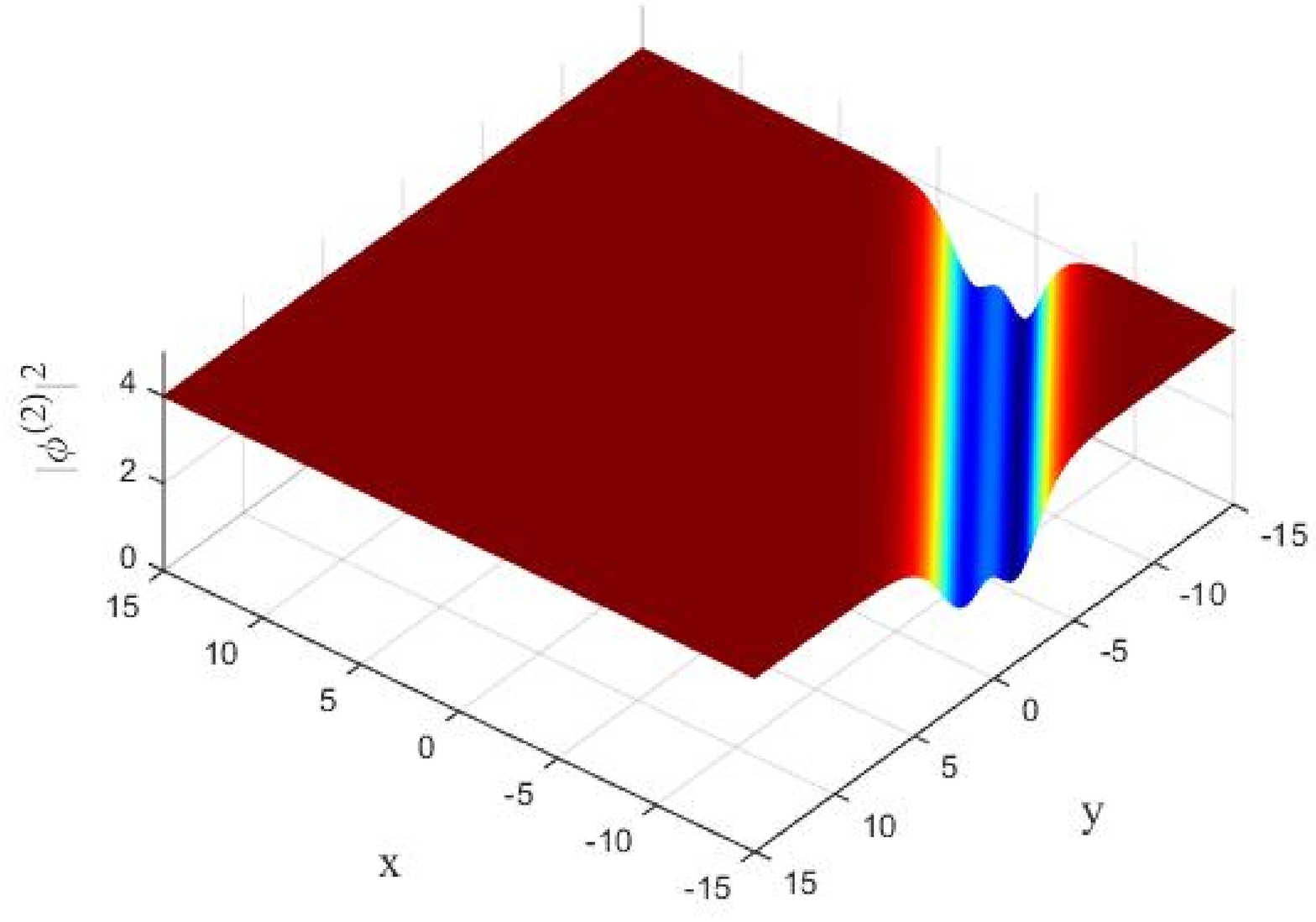}}
\centering
\subfigure[]{\includegraphics[height=0.24\textwidth]{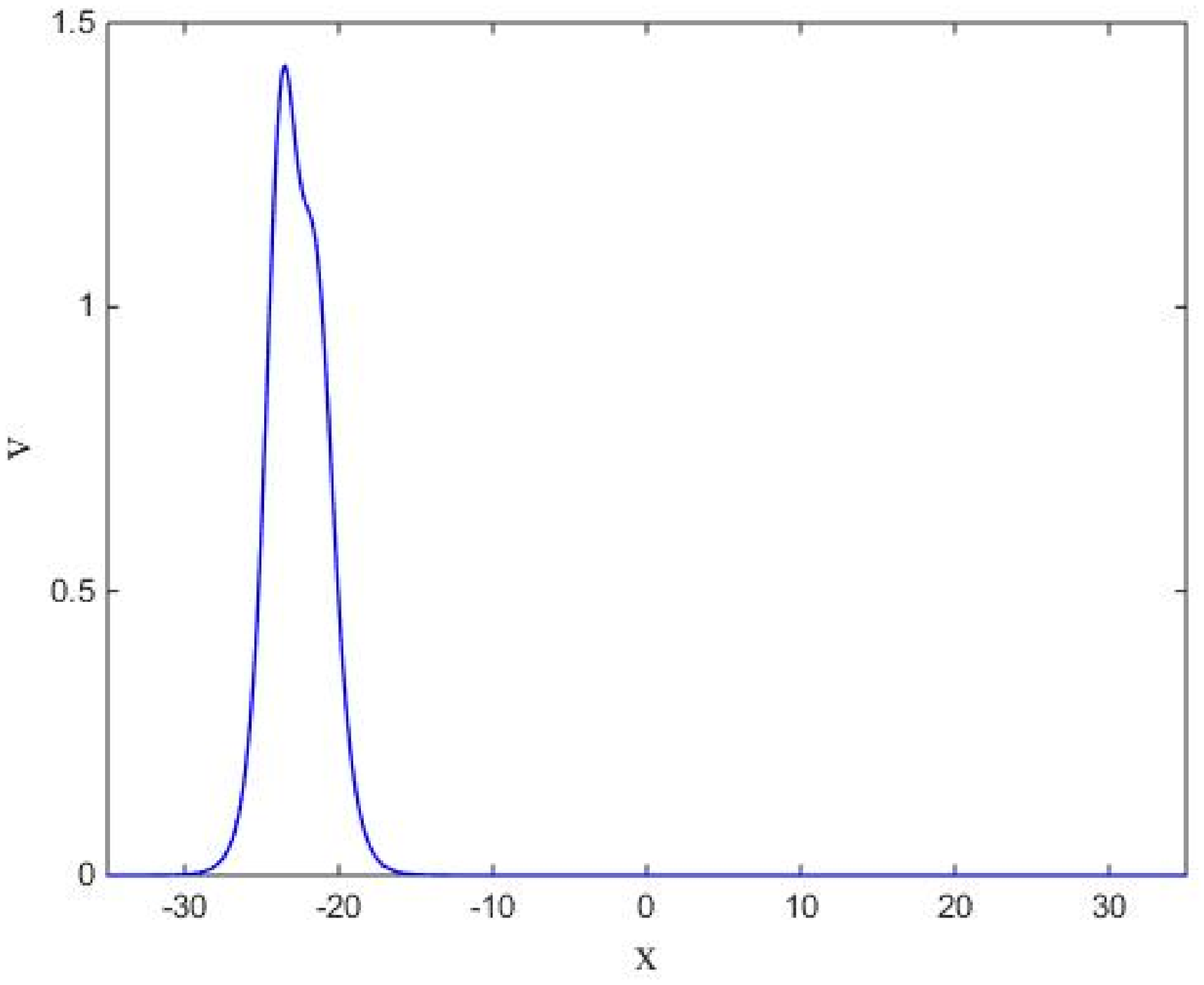}}
\caption{\small (color online) (a)~(b):The moving two dark-dark soliton bound states at time $t=-8$ in the two short wave components; (c) the profile of two bright-bright soliton bound states with $y=10$ at $t=-8$ in the long wave component.\label{xt8-f-6}  }
\end{figure}

\begin{figure}[H]
\renewcommand{\figurename}{{Fig.}}
\subfigure[]{\includegraphics[height=0.24\textwidth]{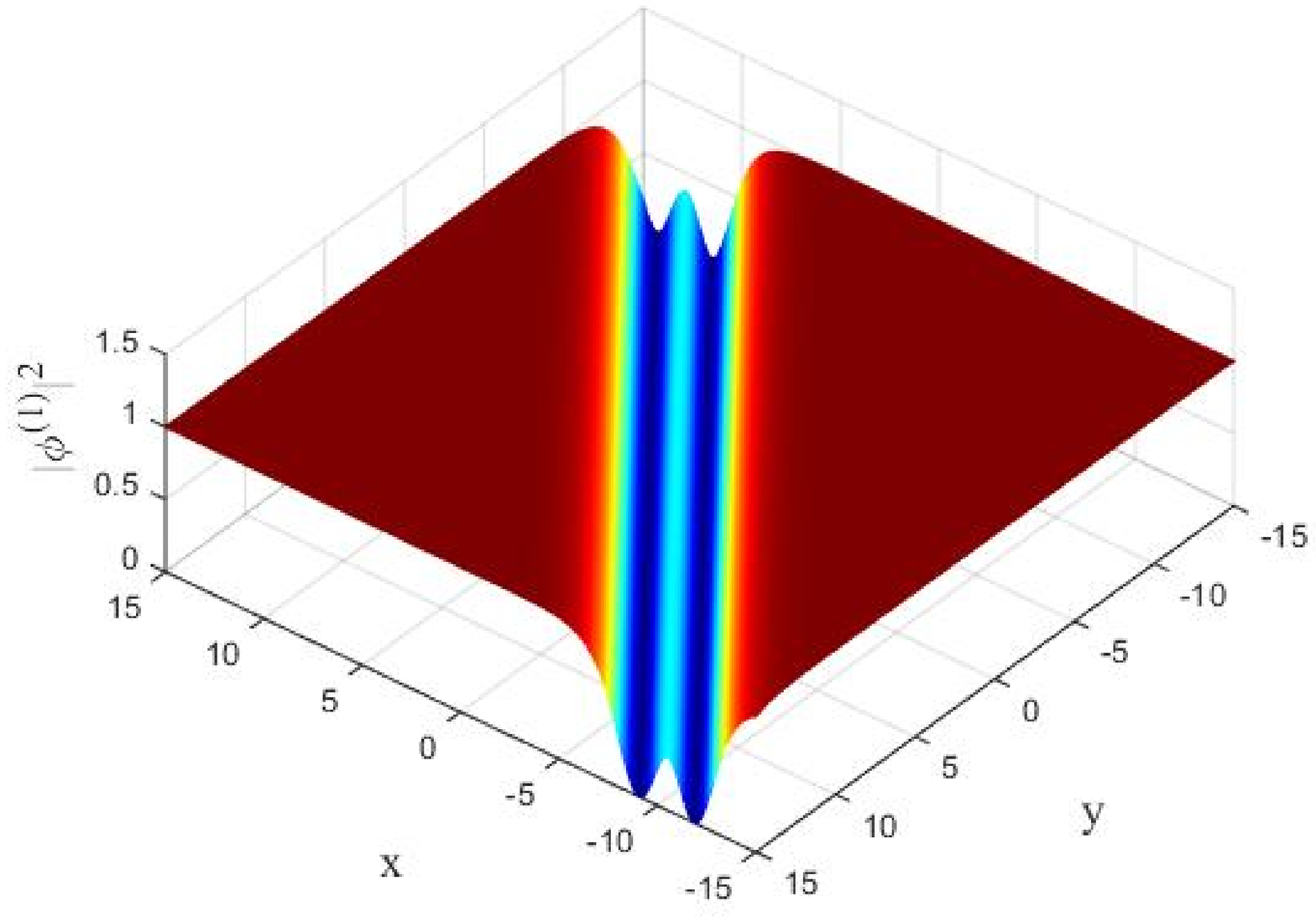}}
\centering
\subfigure[]{\includegraphics[height=0.24\textwidth]{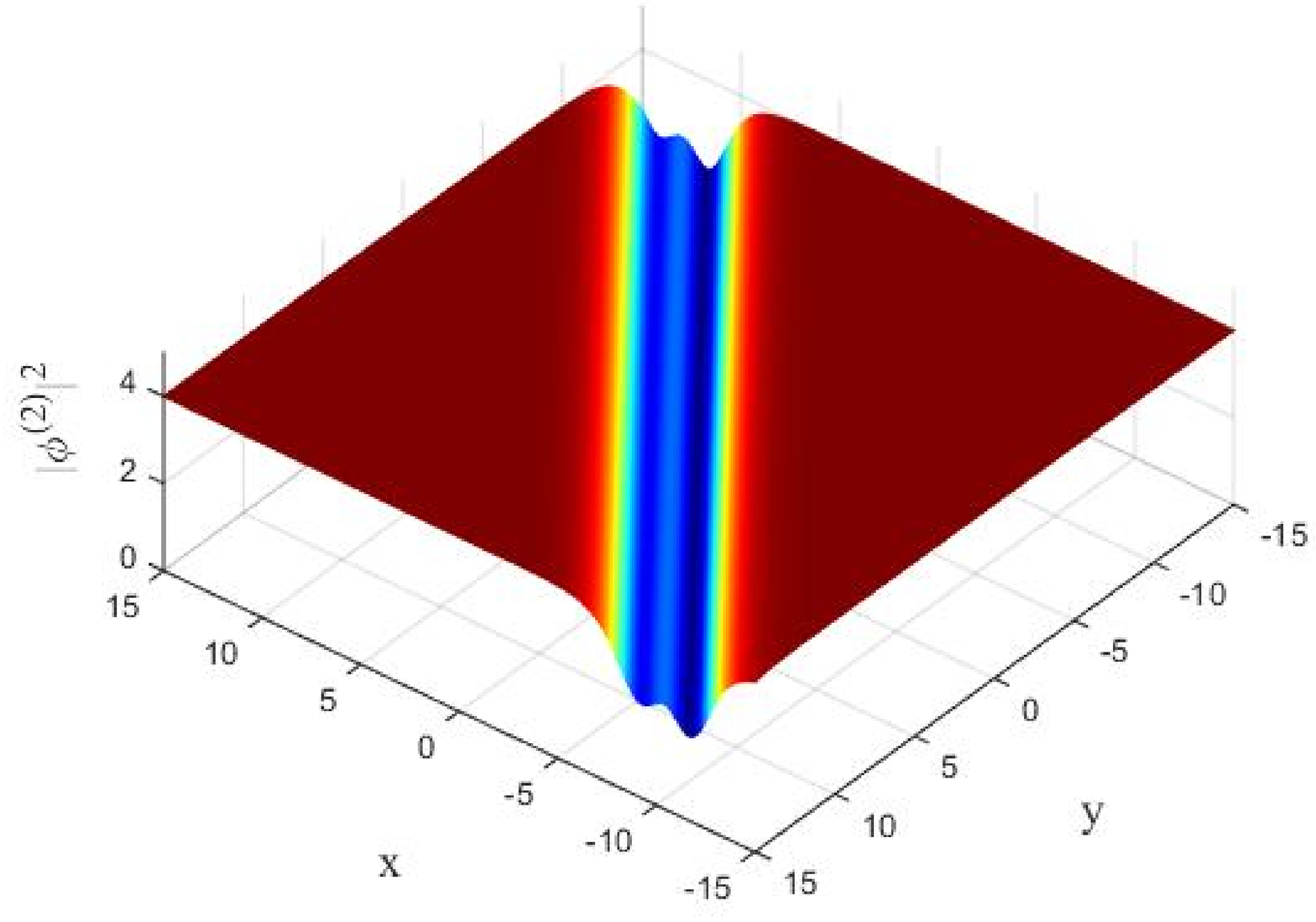}}
\centering
\subfigure[]{\includegraphics[height=0.24\textwidth]{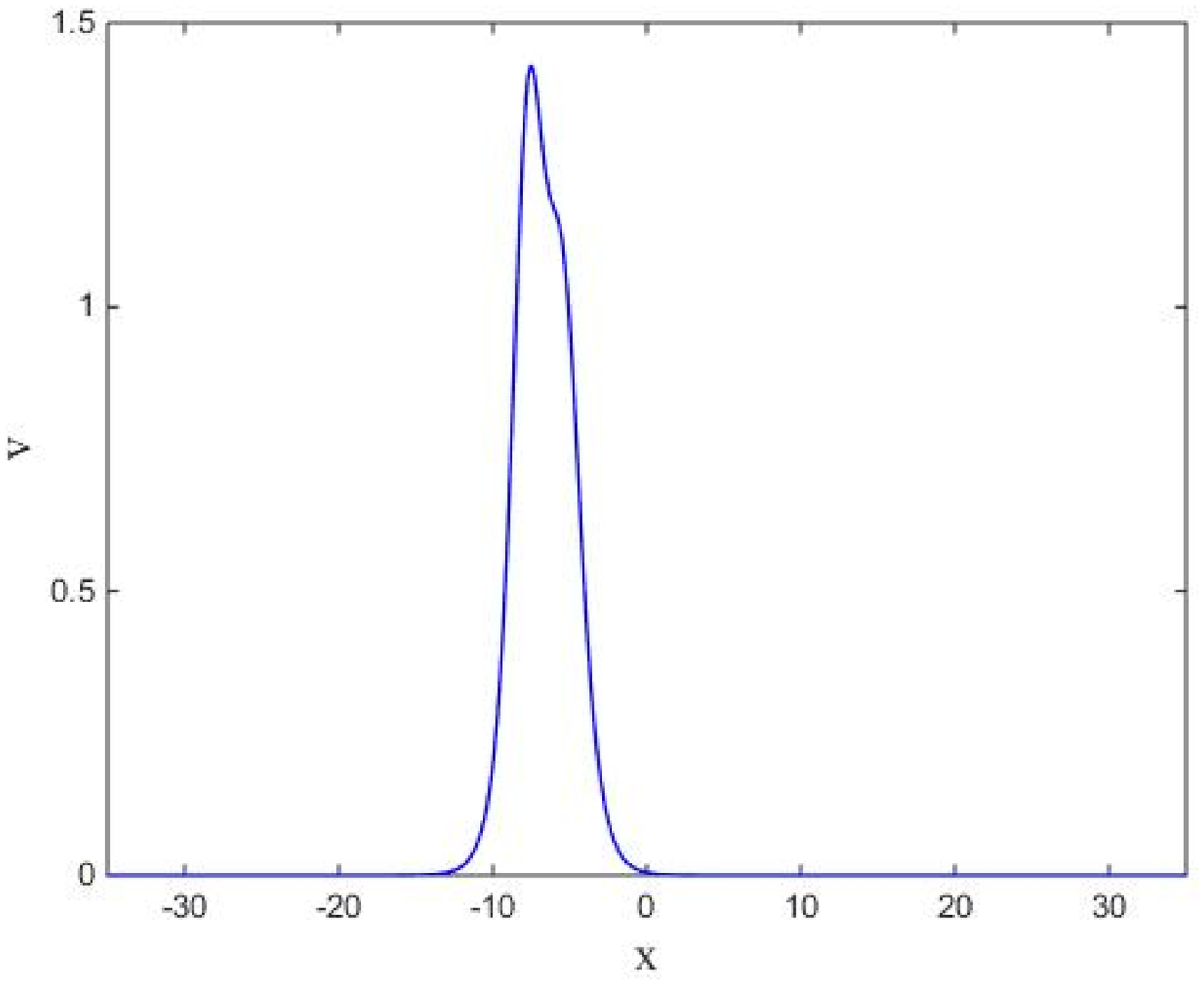}}
\caption{\small (color online)(a)~(b):The moving two dark-dark soliton bound states at time $t=0$ in the two short wave components; (c) the profile of two bright-bright soliton bound states with $y=10$ at $t=0$ in the long wave component.\label{xt8-f-7}  }
\end{figure}

\begin{figure}[H]
\renewcommand{\figurename}{{Fig.}}
\subfigure[]{\includegraphics[height=0.24\textwidth]{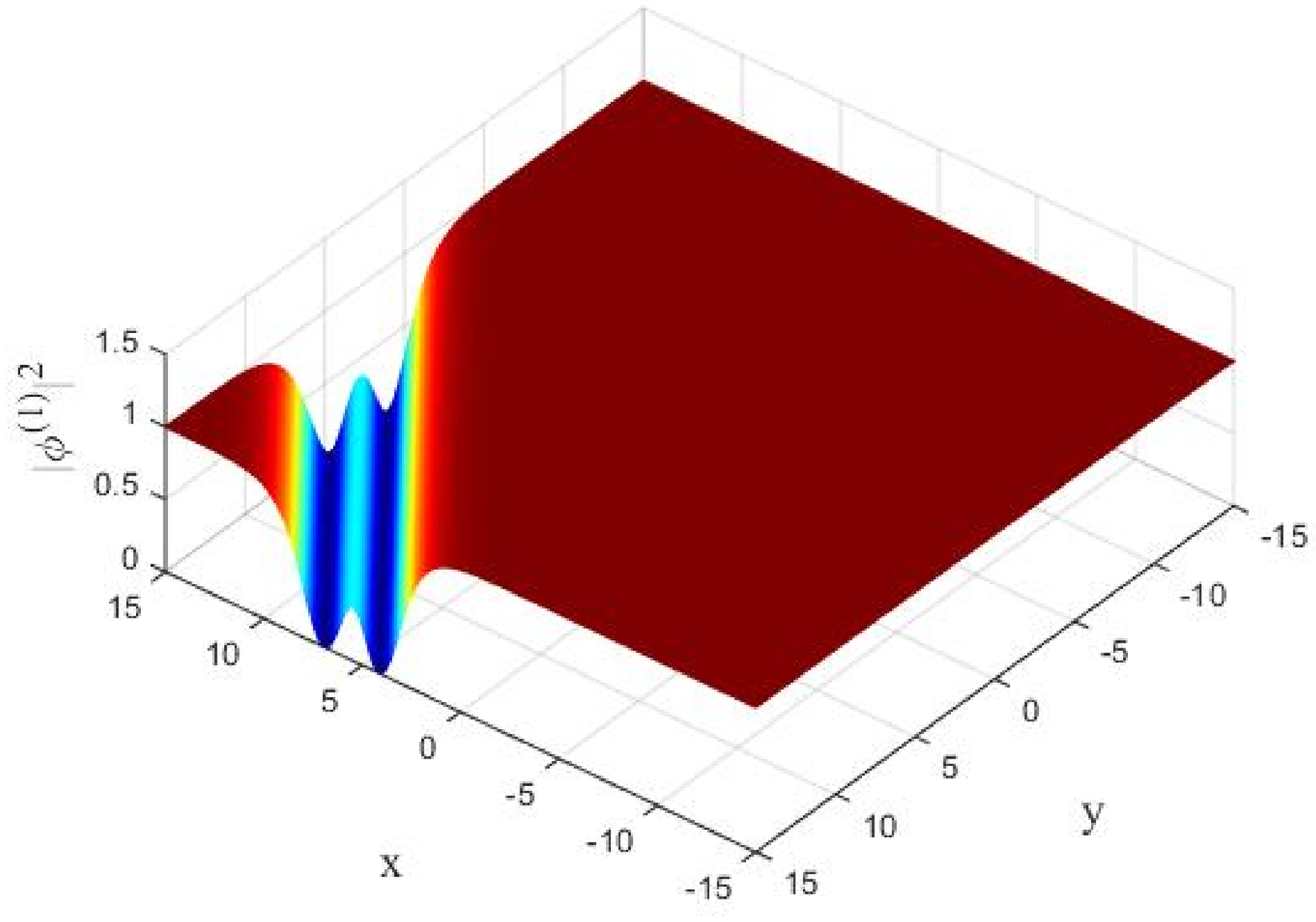}}
\centering
\subfigure[]{\includegraphics[height=0.24\textwidth]{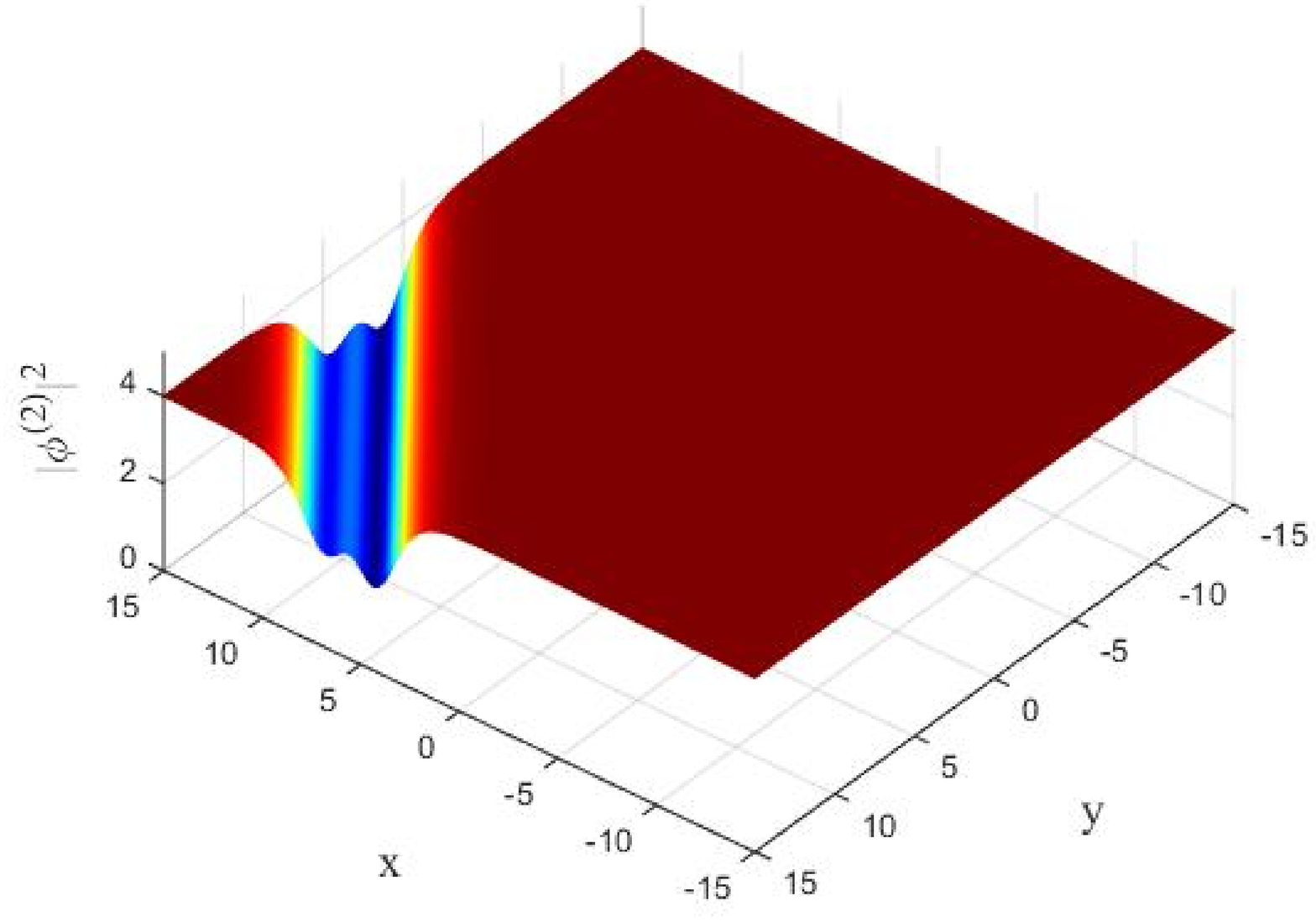}}
\centering
\subfigure[]{\includegraphics[height=0.24\textwidth]{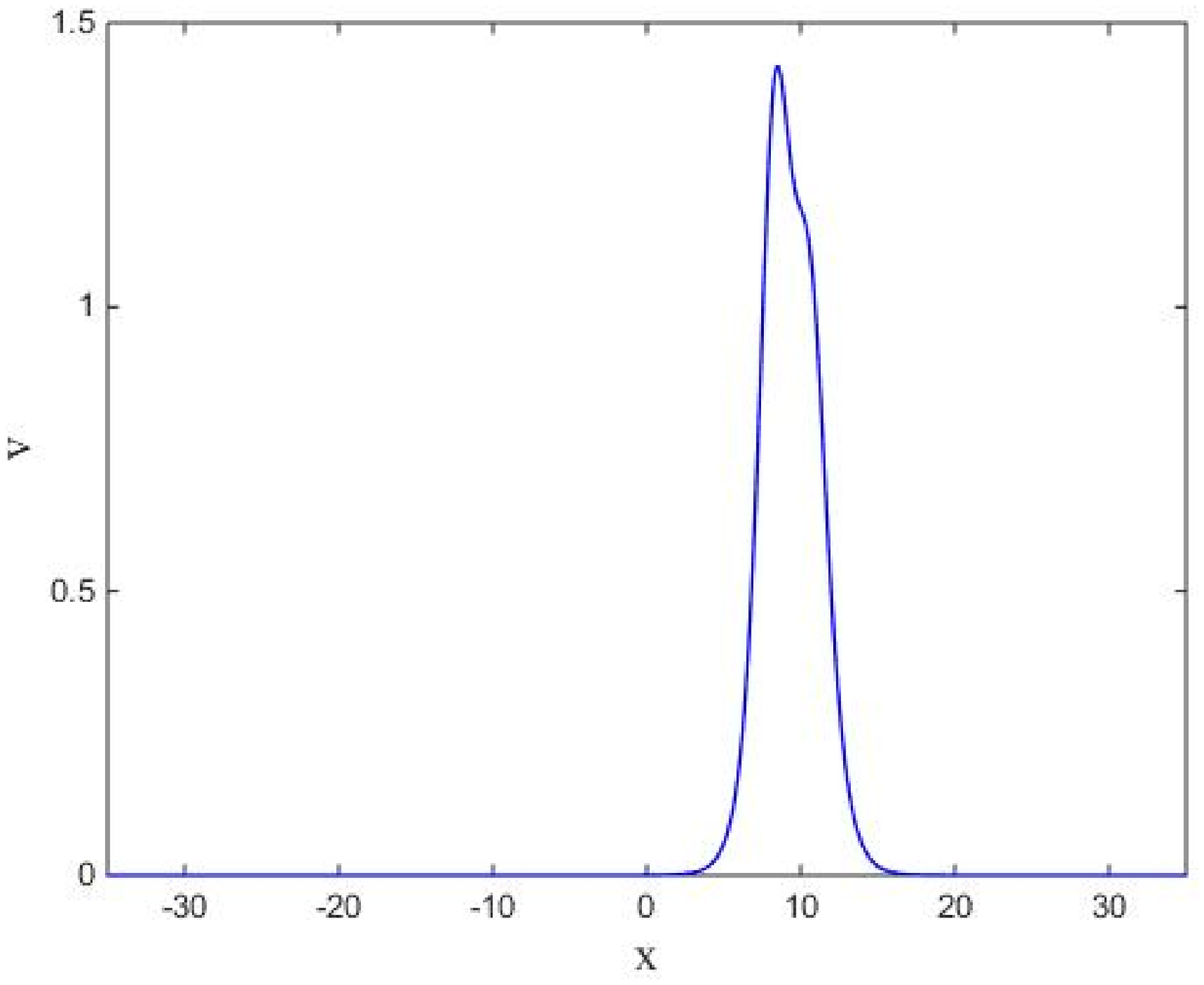}}
\caption{\small (color online) (a)~(b):The moving two dark-dark soliton bound states at time $t=8$ in the two short wave components; (c) the profile of two bright-bright soliton bound states with $y=10$ at $t=8$ in the long wave component.\label{xt8-f-8}  }
\end{figure}

Assuming that the moving bound states can exist up to three dark solitons, and we have $n_1=n_2=n_3$, $g(m_1)=g(m_2)=g(m_3)~(m_1,m_2,m_3>0)$. Form the expression of $g(x)$, we can find that $g(x)$ is the equation
with one independent variable $x$ whose degree is four, and the equation $g(x)=c$ has two positive roots at most. Based on the above facts, it is not possible that $g(m_1)=g(m_2)=g(m_3)$ with $m_1,m_2,m_3>0$. A conclusion can be  made that  only two-soliton bound state exists in the moving case. Additionally, the figures of the moving two-dark-dark solitons are shown in Figs. (\ref{xt8-f-6})-(\ref{xt8-f-8}) at three different times with the parameters chosen by
\begin{eqnarray*}
&&k_1=1, k_2=2, m_1=1, n_1=1,m_2=\frac{5}{52}\sqrt{39},n_2=1, \alpha_1=1,\\
&&\alpha_2=\frac{1}{4}, \sigma_1=1,\sigma_2=-1,\beta_1=0, \beta_2=0, \xi_{10R}=0, \xi_{20R}=0.
\end{eqnarray*}
We find that it is not easy to observe the moving two bright-bright solitons in the long wave component $v$, and we give the corresponding two-dimensional diagrams at $y=10$  in Figs. (\ref{xt8-f-6})(c)-(\ref{xt8-f-8})(c).

 For the two soliton bound states (stationary and moving cases), it is interesting that the two short wave components $\phi^{(1)}$ and $\phi^{(2)}$ have some phase shifts but the long wave component $v$ undergo no phae shift with $x$ and $y$ varying from $-\infty$ to $+\infty$. Additionally, the sum of the  phase shifts of each short wave
component are equal to the sum of the individual ones of the two dark solitons. whereas the phase shift of the long wave component is still zero. Setting that $H_j^{(1)}=e^{2i\psi_j^{(1)}},~H_j^{(2)}=e^{i\psi_j^{(2)}}$, then the phase shifts of the corresponding components can be written as  $\phi^{(1)}_{phase~shift}=2\psi_1^{(1)}+2\psi_2^{(1)}$,~$\phi^{(2)}_{phase ~ shift}=2\psi_1^{(2)}+2\psi_2^{(2)}$ and $v_{phase ~shift}=0$.

\section{$N$-dark soliton solutions for the multi-component generalization}
If the isolated waves localized in small parts  of space are more than three, the Maccari system (\ref{xt-8-1})-(\ref{xt-8-2}) should be generalized to multi-component case. In fact, the previous results for the three-component case can be extended to the ($N+1$)-component coupled Maccari system (\ref{xt-8-4})-(\ref{xt-8-5}). Furthermore, the multi-dark soliton solutions for the ($N+1$)-component Maccari system can be generated from the reduction of single KP hierarchy consisted of $N$ copies of shifted singular points. Similar with the processes in the three-component coupled Maccari system (\ref{xt-8-1})-(\ref{xt-8-2}), we can directly calculate the $N$-dark soliton solutions for the ($N+1$)-component coupled Maccari system (\ref{xt-8-4})-(\ref{xt-8-5}). Here, we only give the main results and omit some complicated calculation procedures.

The ($N+1$)-component coupled Maccari system (\ref{xt-8-4})-(\ref{xt-8-5}) consisting of $N$ short wave components and single long wave component can be transformed into the following bilinear forms
\begin{eqnarray}
&&[D_x^2+i(D_t+2\alpha_lD_x)]h^{(l)}\cdot f=0~(1\leq l \leq N),\\
&&D_xD_yf\cdot f=\sum_{l=1}^{N}{\sigma_lk_l^2(h^{(l)}h^{(l)*}-f^2)},
\end{eqnarray}
through these dependent variable transformations
\begin{equation}
\phi^{(l)}=k_le^{i\theta_l}\frac{h^{(l)}}{f}~(1\leq l\leq N),\quad v=2(lnf)_{xx},
\end{equation}
where $h^{(l)}~(1\leq l\leq N)$ are the complex functions of $x,y,t$, $f$ is the real function of $x,y,t$. Besides, $\theta_l=\alpha_lx-\alpha_l^2t+\beta_l(y)$, $k_l,\alpha_l$ are arbitrary real constants and $\beta_l(y)$ are the arbitrary functions of $y$.

Similar to the reductions in Sec.2, one can construct the $N$-dark soliton solutions in Gram type determinant
forms for the multi-component Maccari system (\ref{xt-8-4})-(\ref{xt-8-5}) as follows
\begin{eqnarray}
&&\phi^{(l)}=k_le^{i\theta_l}\frac{h^{(l)}}{f},\quad v=2(lnf)_{xx},\\
&&f=\left|\delta_{ij}+\frac{1}{p_i+p_j^{*}}e^{\xi_i+\xi_j^{*}} \right|_{N\times N},\\
&&h^{(l)}=\left|\delta_{ij}-\frac{p_i-i\alpha_l}{p_j^{*}+i\alpha_l}\frac{1}{p_i+p_j^{*}} \right|_{N\times N}~(1\leq l\leq N),
\end{eqnarray}
where $\xi_i=p_ix+ip_i^{2}-\frac{1}{2}\sum_{l=1}^{N}\frac{\sigma_lk_l^2}{p_i-i\alpha_l}y+\xi_{i0}$, $\theta_l=\alpha_lx-\alpha_l^2t+\beta_l(y)$, ~$\beta_l(y)$ are the arbitrary functions of $y$, $p_i,\xi_{i0}$ are complex constants and $k_l,\alpha_l$ are real constants, $\delta_{ij}$ is the Kronecker symbol.

\section{Conclusions}
Utilizing the KP hierarchy reduction method, the $N$-dark-dark soliton solutions for the three-component coupled Maccari system (\ref{xt-8-1})-(\ref{xt-8-2}) consisted  of two short wave components and one long wave component are given in Gram determinant forms. There exist some connections between the researched Maccari system and some equations in the KP hierarchy. Here, the related equations are the two-dimensional Toda lattice (Eqs. (\ref{xt-8-6}) and (\ref{xt-8-6-2})) in KP hierarchy and  the lowest-degree bilinear equation (Eqs. (\ref{xt-8-6-1}) and (\ref{xt-8-7})) in the 1st modified KP hierarchy. Based on the above facts, the $N$-dark solitons for the Maccari system (\ref{xt-8-1})-(\ref{xt-8-2}) can be constructed by the complex conjugate reductions and  the independent variable transformations in the KP hierarchy. Similar to the $N$-dark-dark soliton in the three-component case, the $N$ dark solitons for the $(N+1)$-component Maccari system including $N$ short wave components and one long wave component can also be generated. This kind of dark soliton in two-dimensional multi-component system has also been reported for the  YO system \cite{8-21} and Mel'nikov system \cite{8-22}.

In this paper, the dynamics of the dark solitons for the three-component coupled Maccari system are discussed in detail. For one-dark-dark solitons, the degrees of the darkness can be controlled by the parameter $\alpha_j~(j=1,2)$ and they are classified as two types ( degenerate and non-degenerate cases).  From the asymptotic analysis, it is shown the amplitude and velocity of each soliton in the two-soliton solutions maintain unchange during the interaction, then the collisions in the three components are all elastic. The long wave component $v$  always possesses bright soliton and the two bright soliton interactions have O-type and P-type. Additionally, the two dark-dark solitions bound states are discussed in stationary and moving cases. For the stationary bound states, the bound states can exist up to arbitrary order. However, only two-soliton bound state exists in the moving case. There are two types (oblique and parallel) in the stationary bound states. The oblique case can be generated for all possible combinations of nonlinearity coefficients. Nevertheless, the parallel case is only possible when nonlinearity coefficients take opposite signs.

Recently, one optimal control method for dark solitons  were reported in \cite{8-43}. We expect that the $N$ dark solitons obtained in this article will be verified and controlled in the physical experiments in the future.
\section{Acknowledgements}
We would like to express our sincere thanks to other members of our discussion group for their
valuable comments. The project is supported by the National Natural Science Foundation of China (Nos. 11675054, 11435005), Global Change Research Program of China (No. 2015CB953904), Shanghai Collaborative Innovation Center of Trustworthy Software for Internet of Things (No. ZF1213).


\end{document}